\documentclass{aa}  

\usepackage{romannum}
\usepackage{graphicx}
\usepackage{subcaption}
\usepackage{float}
\usepackage[varg]{txfonts}
\usepackage{natbib}         
\bibpunct{(}{)}{;}{a}{}{,} 
\usepackage{bm}				
\usepackage[colorlinks=true,linkcolor=blue,citecolor=blue,urlcolor=blue]{hyperref}		
\usepackage[capitalise]{cleveref}		
\numberwithin{equation}{section}	
\usepackage{xcolor} 
\usepackage{comment}
\usepackage{makecell}

\usepackage[normalem]{ulem}  

\newcommand{\der}{\text{d}} 
\newcommand{\iu}{{\text{i}\mkern1mu}} 
\newcommand{\conj}[1]{\overline{#1}} 
\newcommand{\expv}[1]{\left\langle #1 \right\rangle} 

\begin{document} 
\pagenumbering{arabic}
\setcitestyle{citesep={,}}  

   \title{Scintillometry of Fast Radio Bursts:}
   \subtitle{Resolution effects in two-screen models}

   \author{Sachin Pradeep E. T.
          \inst{1},
          Tim Sprenger\inst{1}, Olaf Wucknitz\inst{1}, Robert A. Main\inst{2,3,1} and Laura G. Spitler\inst{1}
          }

   \institute{Max-Planck-Institut für Radioastronomie,
              Auf dem Hügel 69, D-53121 Bonn, Germany\\
              \email{sachinpradeepet@gmail.com; tsprenger@mpifr-bonn.mpg.de}
         \and
             Department of Physics, McGill University, 3600 rue University, Montr\'{e}al, QC H3A 2T8, Canada
         \and
             Trottier Space Institute, McGill University, 3550 rue University, Montr\'{e}al, QC H3A 2A7, Canada
             }

   \date{Received \today} 

  \abstract
    {Fast Radio Bursts (FRBs) exhibit scintillation and scattering, often attributed to interactions with plasma screens in the Milky Way and the host galaxy. When these two screens appear "point-like" to each other, two scales of scintillation can be observed with sufficient frequency resolution. A screen perceives a second screen as extended or resolved when the angular size of the latter is smaller than the angular resolution of the former. We define the ratio of these two as the Resolution Power (RP). Previous observational studies have argued that, in the resolving regime, scintillations disappear, assuming that a screen resolving another screen is equivalent to a screen resolving an incoherent emission region. In this theoretical and simulation-based study of resolving effects in two-screen scenarios, we argue that resolving quenches only the relatively broad-scale scintillation and that this quenching is a gradual process. We present qualitative and quantitative predictions for dynamic spectra, spectral autocorrelation functions (ACF), and modulation indices in resolved and unresolved regimes of two-screen systems. We show that the spectral ACF of a two-screen system has a product term in addition to the sum of individual screen contributions, causing the total modulation index to rise to \(\sqrt{3}\) in the unresolved regime. To aid in discovering resolving systems, we also present observable trends in multi-frequency observations of a screen resolving another screen or incoherent emission. Additionally we introduce a new formula to estimate the distance between the FRB and the screen in its host galaxy. We also show that this formula, like previous ones in the literature, is only applicable to screens that are two-dimensional in the plane of the sky.}


   \maketitle

\section{Introduction}
\label{Sec:Intro}

Fast Radio Bursts (FRBs) are extra-galactic single pulses of unknown origin that range from micro- to milliseconds in duration. Since their first detection in 2007 \citep{Lorimer2007Sci...318..777L}, almost 1000 FRBs have been detected, mostly by CHIME-FRB  at 400-800 MHz at high Galactic latitudes (\citealt{Amiri_2021}). The extra-galactic origin of FRBs has been confirmed through precise interferometric localization of the bursts and host galaxy associations \citep[e.g.][]{Chatterji2017Natur.541...58C, tendulkar2017host,2019Sci...365..565B}. The redshifts of localized sources currently extend up to  $z\lesssim1$ \citep{2023Sci...382..294R}.

One or more plasma lenses located anywhere along the burst's line of sight can induce multi-path propagation. As a result, an image-domain observer sees the lens as an angularly broadened source, while a time-domain observer sees a delayed or temporally broadened pulse, often manifesting as a scattering tail, or an interference pattern in frequency, known as scintillation. The scattering tail is caused by group delays, while scintillation is caused by interference due to different phase delays of the radio waves. These effects are quantified as the scale of angular broadening ($\theta_L$), scattering time ($\tau_s$) and the scintillation bandwidth ($\nu_s$), respectively. Moreover, $\tau_s$ and $\nu_s$ produced by the same screen are related through the Fourier uncertainty principle (\(2 \pi \tau_s \nu_s = C\), where C ranges between 0.5 and 2 \citep{rickett_lambert_1999}), and therefore any screen can be characterized by either of these quantities. In practice, to observe and characterize the scattering time scale \(\tau_s\), the scattering delays must exceed the pulse width, and the observational time resolution needs to be finer than the scattering time scale. Similarly, to observe scintillation in frequency, the frequency resolution of the data should be finer than the scintillation bandwidth \(\nu_s\). A plasma lens is often approximated as a thin screen \citep{Williamson1972MNRAS.157...55W}, because they are much thinner than the distances between the source, screens, and observer. In this work, the size of the scattering screen is defined as the extent of the scatter-broadened source, which is different from the physical size of the plasma structure giving rise to the screen. The full physical sizes of plasma structures can be much larger and are difficult to probe.

An FRB traverses several ionized media between its origin and our telescope, including the interstellar medium (ISM) and halo of the host galaxy, the intergalactic medium (IGM), and the ISM and halo of the Milky Way (MW). In addition, some FRBs may originate in a dense circum-burst medium (CBM) or encounter the ISM or halo of an intervening galaxy with the probability of an intersection increasing with distance. 
In principle, any of these media could introduce scattering and scintillation in a FRB. 
In many FRBs both pulse broadening and frequency scintillations are observed, but the scales violate the  Fourier uncertainty principle condition described above  \citep[e.g.][]{masui2015dense,Farah2018MNRAS.478.1209F,sammons2023two}, implying that scattering screens in two different media are contributing to the observed scattering effects. Thus, it is reasonable to assume that many FRBs encounter two or more plasma screens. 
The MW scattering budget, which can be estimated from pulsar observations, is of order $\nu_s \sim 1-10$~MHz ($\tau_s \sim 100 - 10\,$ns) at a frequency of 1~GHz for Galactic latitutes $\gtrsim 30$ degrees \citep{cordes2003ne2001}, and the observed frequency scale of scintillation in FRBs often matches the MW expectation  \citep[e.g.][]{masui2015dense,Schoen_2021,main2022scintillation}. \citet{Ocker2021ApJ...911..102O} showed that scattering in the MW halo is at least an order of magnitude lower than for the thin and thick disk of the MW, and presumably the same can be assumed for the host galaxy halo. A theoretical prediction for scattering in the IGM of $<$1~ms at 300 MHz \citep{macquart2013temporal}, as well as an observational constraint from the intersection with a filamentary structure \citep{2024arXiv241007307S}, suggest that the IGM contributes negligably to FRB scattering. 
The likelihood of an FRB intersecting the ISM of a foreground galaxy, and thus experiencing extreme scattering, is less than 5\% for \(z<1.5\) \citep{macquart2013temporal}. 
On the other hand, the probability of an FRB's sight line intersecting the circumgalactic medium of a foreground galaxy is significantly higher; an FRB will intersect, on average, one or more halos of mass $M_{h} > 10^{11} M_{\sun}$ and $M_{h} > 10^{13}  M_{\sun}$ (where $M_{\sun}$ is a solar mass) above a redshift of $z>0.1$ and $z>1$, respectively \citep{2014ApJ...780L..33M}. 

Therefore, the pulse broadening observed in FRBs originates in the CBM, the ISM of the host galaxy, or an intervening halo, or the MW ISM, and which of these dominate likely depends on the source and its line of sight. A large sample of scattered FRBs discovered by CHIME/FRB was best described by a population model that included the first three of these constituents \citep{chawla2022modeling}. 
Temporal variations in $\tau_s$ seen from FRB~20190520B strongly argue for pulse broadening originating in the CBM \citep{2023Ocker}, but this source is known to reside in an extreme local environment \citep{2023Sci...380..599A} and may not be representative of the population as a whole. 
Disentangling whether the scattering originates in the host ISM or a foreground halo requires identifying possible foreground galaxies from optical images of the FRB field.  
For example, the scattering in FRB~20190608B likely originates in the host galaxy despite the presence of an intervening halo \citep{2021ApJ...922..173C, 2020ApJ...901..134S, sammons2023two}, and an upper limit to scattering in FRB~20181112 suggests that the intervening halo introduced little pulse broadening \citep{2019Sci...366..231P}.
On the other hand,  \citet{2024arXiv240514182F} argue that the large pulse broadening seen in FRB~20221219A could originate in a dense cloud in one of the two intervening halos along the line of sight, and \citet{2020MNRAS.499.4716C} show that scattering in the halo of M33 introduced frequency scintillations on a scale broader than the Milky Way ISM. 

As an interference phenomenon, scintillation requires coherent radiation, which means that incoming waves have a stable phase relation to each other. However, most emssion mechanisms produce radiation whose phase rapidly and randomly varies such that they are not \emph{temporally} coherent. Instead, scintillation is caused by \emph{spatial} coherence which means that radiation incoming along one path has a stable phase relation to radiation incoming along other paths of propagation. Spatial coherence follows from the source being point-like since only one starting point of each path leads to all emitted radiation effectively getting summed into a single spherical wave. Thus, the phase of observed waves can only differ by the phase delay acquired on the way. A source is considered extended instead of point-like when emission from different points on the source leads to geometric phase changes, with respect to emission from other points, that are large enough to cause destructive interference. In this case, instead of being effectively combined into a single point source, there are independent patches of emission that are not coherent with respect to each other. Now, we call the source resolved with the size of these patches being a measure of resolution. The loss of coherence results in suppressed scintillation. This process is referred to as quenching. A common analogy visible to the human eye is that stars scintillate due to atmospheric fluctuations, while planets due to their larger angular size do not. 

\cite{masui2015dense} proposed a reasoning where scattering from a screen in the host galaxy broadens a FRB such that it can be resolved by the Milky Way screen, which would quench the scintillation corresponding to the Milky Way screen. This argument has been used to place constraints on the distance between the FRB and the screen closer to it, and thus on the immediate source environments, progenitors, and emission mechanisms (see e.g., \citealp{masui2015dense, Farah2018MNRAS.478.1209F,cordes2019fast, 2022ApJ...931...87O, main2022scintillation, sammons2023two, 2025Natur.637...48N}). However, it is not strictly correct that a point source, scatter-broadened by a screen, becomes equal to an extended physical source because patches on a scattering screen still have stable phase relationships to each other in contrast to patches on the physical source. In this work, we derive observables from a two-screen model and investigate the phenomenon of quenching in numerical simulations within this model. Mathematically, the order of the screens is irrelevant, so the question needs to be answered of whose screen's corresponding scintillation disappears. Previously, the observational studies of two-screen scattering in FRBs relied heavily on existing single-screen theory, indicating a need for a systematic theoretical and numerical study to support future research and understand earlier results. In this study, we position one scattering screen in the ISM of the host galaxy and another within the Milky Way's ISM, but the results are also valid for a screen in an intervening halo. Although the focus is on FRBs, this study applies to multi-screen effects on all radio pulses, including those from pulsars.

In this work, we discuss in detail how observables such as burst morphology, dynamic spectra, modulation index, and spectral auto-correlation function are affected as they encounter two plasma screens. This paper is structured as follows: In \cref{Sec:CosmoScatteringTheory}, we expand the local framework of scintillation theory to develop a cosmological framework for studying multi-plane scattering. Next, in \cref{Sec:Models}, we explain how scattering screens are modeled and discuss the two-screen resolving argument. In \cref{Sec: Observables}, we derive the theoretical forms of the studied observables. In \cref{Sec: FRB_Scintillator}, we present {\tt FRB\_scintillator}, a new two-screen simulation tool, along with our analysis methodology. The results of our simulations are detailed in \cref{Sec: Simulation Results}, followed by an in-depth discussion of their implications and potential applications in \cref{Sec: Discussion}.

\section{Cosmological two-screen geometric delays}
\label{Sec:CosmoScatteringTheory}

\begin{figure}
    \centering
    \includegraphics[width=\columnwidth]{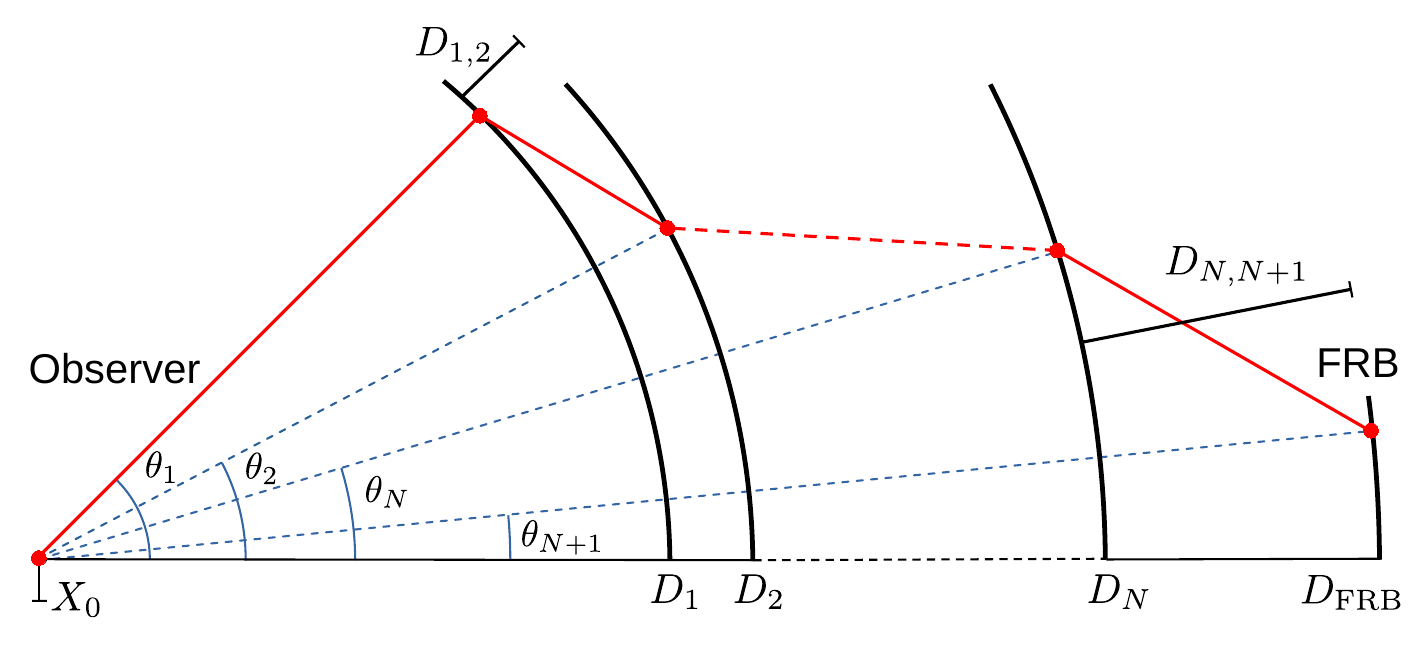}
    \caption{Definitions of angles and distances in a system where a ray originating from an FRB is scattered $N$ times. Figure adapted from \citet{2023MNRAS.520.2995F}.}
    \label{fig:CosmologicalAngles_illustration}
\end{figure}

\begin{figure}
    \centering
    \includegraphics[width=\columnwidth]{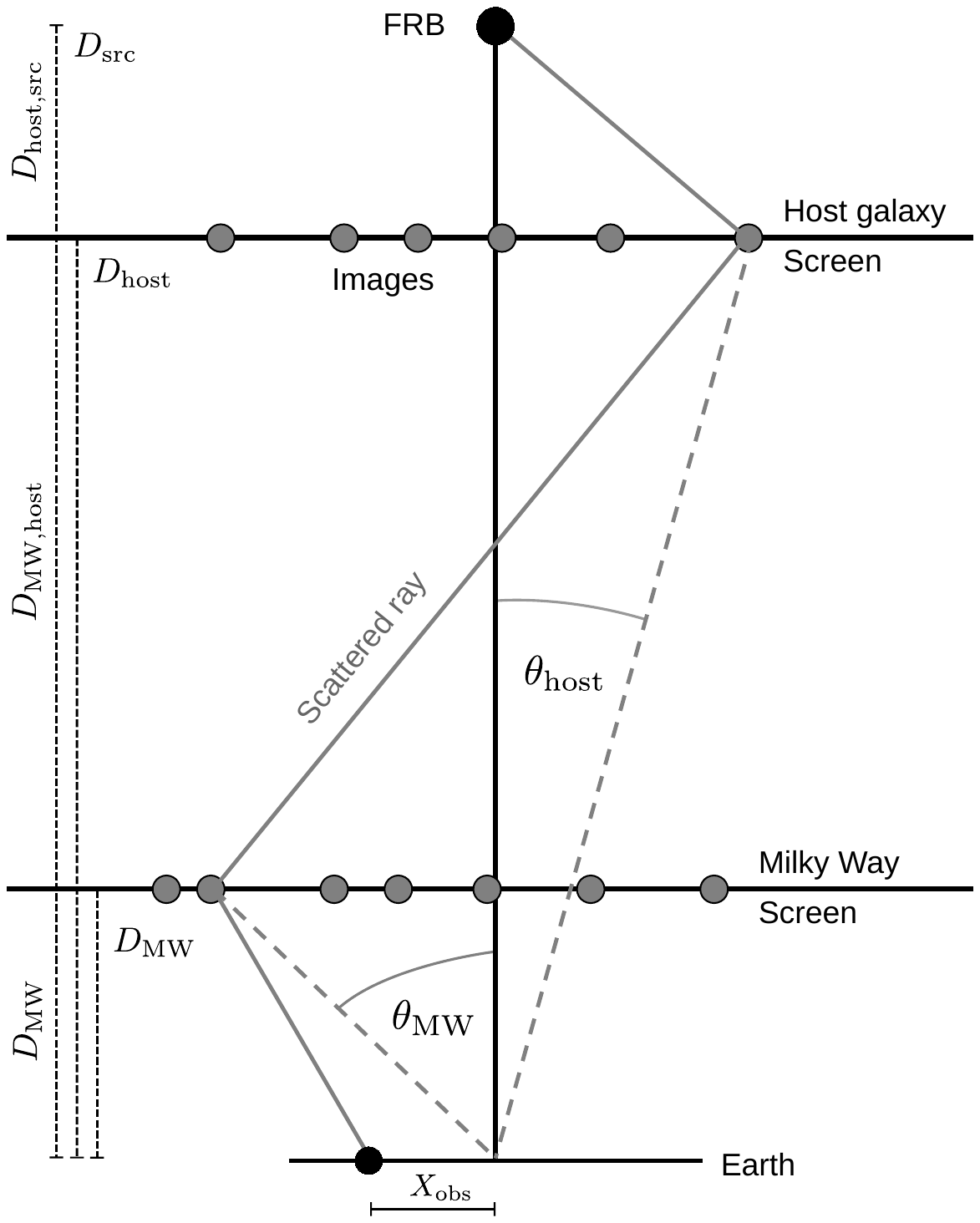}
    \vspace{-1cm}
    \caption{Illustration of a ray originating from a fast radio burst and scattered by two screens at points refered to as images. The distances of the MW, host galaxy, and source as well as the relative distances between them are introduced. The angular positions of images are defined with respect to an optical axis while the position of the observer might be offset by $X_\text{obs}$.}
    \label{fig:2screen_illustration}
\end{figure}

The study of pulsar scintillation has led to a picture of localized scatterers on thin scattering screens, whose positions do not evolve with frequency and time within a single observation. The same plasma structures that cause pulsar scintillation in the Milky Way will also cause scintillation in FRBs, and presumably similar structures exist in host galaxies of FRBs. These scatterers are of unknown physical nature and cause separated point-like images of the FRB at their position. While this picture is certainly an approximation and may not be applicable in all cases, it has proven to be very helpful in modeling persistent structures in scintillation arcs \citep{walker2004interpretation,2005ApJ...619L.171H,Cordes_2006} and is numerically tractable. Extensions for multiple screens have been derived by \citet{2022MNRAS.515.6198S} and \citet{2023ApJ...950..109Z}. For an arbitrary number of screens the delay of a scattered ray relative to a hypothetical unscattered ray is given by
\begin{equation}
    \tau_\text{astro} = \frac{D_\text{scattered}-D_\text{direct}}{c} \simeq \sum_{n=1}^{N+1} \frac{1}{2c} \, \frac{\left( \bm{X}_n - \bm{X}_{n-1} \right)^2}{D_{n-1,n}} \label{Eq:psr_delay}
\end{equation}
where N is the number of screens and $D_{n-1,n} = D_{n}-D_{n-1}$. $n=0$ denotes the observer and $n=N+1$ denotes the source. This formula follows from simple geometry of small positional shifts $\bm{X}$ of the scatterers relative to an arbitrary point of reference. This description relies on an overall Euclidean geometry, which no longer holds on a cosmological scale, when the expansion of space and spatial curvature have to be taken into account. 

\citet{1992grle.book.....S} formulate the geometric delay as a sum of cumulative delays resulting from differences of the apparent angles of wavefronts as seen from Earth (see \cref{fig:CosmologicalAngles_illustration}):
\begin{equation}
    \tau_\text{cosmo} = \sum_{n=1}^{N} \frac{1+z_n}{2c}\,\frac{D_n D_{n+1}}{D_{n,n+1}}\left(\bm{\theta}_{n+1}-\bm{\theta}_{n}\right) ^2 - \frac{1}{c}\,\bm{X}_0\cdot\bm{\theta}_1
    \label{Eq:frb_delay}
\end{equation}
where $z_n$ is the redshift of screen $n$ and $D$ are now angular diameter distances. This relation holds for an arbitrary global spacetime and can be derived using only the definition of angular diameter distances $D_{a,b}=X_b/\theta$, where $\theta$ is the angular diameter under which an object at position $b$ of physical size $X_b$ appears to an observer at $a$, and a reciprocity relation of \citet{1933PMag...15..761E}. The subtraction of the observer's offset $\bm{X}_0$ multipled by $\bm{\theta}_1/c$ is not present in \citet{1992grle.book.....S} because they assumed a fixed observer. For scintillation however, the motion of the observer can be crucial if a screen is much closer to the observer than to the source. However, all angles $\theta$ are defined with respect to a fixed reference position instead of the potentially changing observer's position:

\begin{equation}
    \bm{X}_n = D_n \bm{\theta}_n \, .
\end{equation}
The delay in \cref{Eq:frb_delay} is written to second order in the coordinates of the source and all scattered image positions. The observer position, however, is only considered linearly in the mixed term with $\bm{\theta}_1$. In reality, there is also a quadratic term (curvature of the wavefront) in $\bm{X}_0$. This term cannot be derived using the angular diameter distances alone, but requires the parallax distance, and thus additional information about the cosmological model. Because this term does not matter for our analysis, we neglect it entirely. 

The delay as given in \cref{Eq:frb_delay} differs from a formulation derived in \citet{macquart2013temporal} that has been employed by a part of the literature on FRB scintillation and scattering. \footnote{
A \emph{convention} used by \citet{macquart2004} in Eq.~(5) and \citet{macquart2013temporal} in Eq.~(1) allows to write these terms as the square of a difference of scaled positions. Unfortunately, this form leads to a mixed term that is inconsistent with the last term in \cref{Eq:frb_delay}, which is actually physically relevant, unless a factor of $1+z_\text{L}$ is added to the denominator in the terms of $\bm{X}$ or $\bm{X'}$ in those equations. Thus, we argue that a correction needs to be applied to Eqs.~(3,11,12,13) in \citet{macquart2013temporal}, so that this factor moves from the denominator to the numerator in Eq.~(15). Furthermore, a factor of $2$ needs to be added in the denominator. This correction affects every study that made use of Eqs.~(15) and (16) in \citet{macquart2013temporal}. We identified \citet{sammons2023two}, \citet{2023RvMP...95c5005Z}, \citet{2023ApJ...946L..18P}, \citet{2020MNRAS.498.4811H}, \citet{2020MNRAS.497.1382Q}, \citet{2018ApJ...865..147Z}, and \citet{2018MNRAS.474..318P} as being affected. \citet{2025Natur.637...48N} make use of the Eq.~(15) but neglect redshift contributions. \citet{2017ApJ...847...19D} and \citet{2016ApJ...832..199X} cite \citet{macquart2013temporal} but do not seem to be affected because they later use formulas consistent with our result.
}

If all redshifts are zero, \cref{Eq:frb_delay} becomes equal to \cref{Eq:psr_delay} up to terms that only include the observer's or the source's position. These terms have no impact on the scintillation because they are constant for all paths and thus represent only an arbitrary constant that can be added to all delays.

Now, we can regard the specific case of two screens of which one resides in the MW and one in the host galaxy which is illustrated in \cref{fig:2screen_illustration}. Hence, there is only one nonzero redshift involved which will simply be denoted as $z$. The notation of the subscripts can now be made more descriptive: 0 is the observer ($_\text{obs}$), 1 is the MW ($_\text{MW}$), 2 is the host galaxy ($_\text{host}$), and 3 is the source ($_\text{src}$). After neglecting terms that do not contain $\bm{\theta}_{\text{MW}}$ or $\bm{\theta}_\text{host}$, \cref{Eq:frb_delay} becomes
\begin{equation}
\begin{split}
    \tau =\, &\frac{1}{2c} \Bigg( \frac{D_{\text{MW}} D_{\text{host}}}{D_{\text{MW,host}}} \left(\bm{\theta}_{\text{MW}} - \bm{\theta}_{\text{host}}\right)^2 + (1 + z) \frac{D_{\text{host}} D_{\text{src}}}{D_{\text{host,src}}} \bm{\theta}_{\text{host}}^2 \Bigg) \\
     &- \frac{1}{c}\,\bm{X}_\text{obs}\cdot\bm{\theta}_\text{MW} - \frac{1+z}{c}\,\frac{D_{\text{host}} D_{\text{src}}}{D_{\text{host,src}}}\,\bm{\theta}_\text{host}\cdot\bm{\theta}_\text{src} \, .
\end{split}
\label{Eq:delay_formula}
\end{equation}

The last two terms of \cref{Eq:delay_formula} describe the offset and, in the time dependent case, the motion of observer and source from the reference position. For a single burst, movement is not important, which is the case regarded in this work. For the case of repeating bursts, the evolution manifests as a Doppler rate whose corresponding derivation can be found in \cref{sec:fD_2scr}.

Setting the initial positions of observer and source to the origins of their respective planes, we eliminate the last two terms in equation \cref{Eq:delay_formula}; the delay can then be described in familiar form by defining effective distances: 
\begin{equation}
    \tau = \,  \frac{D_{\text{eff,MW}}}{2c} \bm{\theta}_{\text{MW}}^2 - \frac{D_{\text{eff,MW}}}{c}\bm{\theta}_{\text{MW}}\cdot\bm{\theta}_{\text{host}} + \frac{D_{\text{eff,host}}}{2c} \bm{\theta}_{\text{host}}^2  \, . \label{Eq:tau_eff}
\end{equation}
This formulation is phenomenological since it absorbs degenerate parameters into independent degrees of freedom. It is also closely following the canonical formulation of the one-screen case. The effective parameters are given by
\begin{subequations}
\begin{align}
    D_{\text{eff,MW}} &= \frac{D_{\text{MW}} D_{\text{host}}}{D_{\text{MW,host}}}  \approx D_\text{MW} \, ,\\
    \begin{split}
        D_{\text{eff,host}} &= (1 + z) \frac{D_{\text{host}} D_{\text{src}}}{D_{\text{host,src}}} + \frac{D_{\text{MW}} D_{\text{host}}}{D_{\text{MW,host}}} \approx \frac{(1 + z) D^2_{\text{src}}}{D_{\text{host,src}}} \, .
    \end{split}
\end{align}
\end{subequations}

\section{Modelling scattering screens}
\label{Sec:Models}

Radio telescopes measure the time-varying electric field of radio waves which contains the intrinsic wave emitted by the source as well as propagation effects described by the delays imposed at each path of propagation.

The effect of scattering (in the ISM or elsewhere) can be described as a response function with which the intrinsic electric field is convolved:
\begin{equation}
    E(t) = (E_{\text{int}} * R)(t) \, .
    \label{Eq:def_E_measured}
\end{equation}
The intrinsic electric field $E_{\text{int}}$ is the unperturbed emission from the source and $R(t)$ is the impulse response function. All possible paths of propagation can be obtained by taking all pairs of angular scattering positions. Positions in the MW screen are sorted by number $m$ and images in the host screen are counted by number $l$. Furthermore, each scatterer imposes a magnification $\mu$ on the radiation passing through, such that each path has a delay $\tau$ given by \cref{Eq:tau_eff} and a complex amplitude $\mu$ that represents the magnification of the voltage: 
\begin{equation}
    R(t) = \sum_{m,l} \mu_m\,\mu_l\, \delta\left( t - \tau(\bm{\theta}_{\text{MW},m},\bm{\theta}_{\text{host},l}) \right) \, .
    \label{Eq:IRF_time}
\end{equation} 

Several approximations were made here whose validity needs to be discussed. The positions of scatterers are frozen in time and frequency. This is only approximately true for short observation times and narrow bandwidths. Nevertheless, this assumption has been successfully employed to model many observations of scintillation. Also, frequency-dependent delays due to different dispersion measures of each path were neglected. This assumption is again justified by the success of purely geometrical models of scattering delays in describing the main features of scintillation. The magnifications of all paths are considered to be independent of time and frequency for the same reasons as their position. Finally, positions and magnifications are also approximated to be independent from the angle of incoming rays scattered by other screens. This is a crucial approximation that was made to simplify the derivations. In future work, more general models that work without this approximation might be needed to interpret observations. 

Our assumptions are only valid deep inside the regime of geometric optics. There is an alternative tradition of using models based on wave optics for pulsars and also FRBs \citep[e.g.][]{Ocker2021ApJ...911..102O}. Although the ranges of applicability of both regimes have been studied theoretically \citep[e.g.][]{2023MNRAS.520.2995F,2023MNRAS.525.2107J}, limited information makes it difficult to decide in practice. The biggest difference in the inferred results is the density and distribution of plasma, which will not be discussed here. Since a cloud of fixed rays is able to approximate a single diffracted beam, we expect our conclusions for the size and distance of scattering screens to still be valid in other regimes to an extent that only future studies can determine.

Within our approximation, it is much more convenient to compute $\tilde{R}(\nu)$ in frequency space, where it becomes
\begin{equation}
    \tilde{R}(\nu) = \sum_{m,l} \mu_m\,\mu_l\, \text{e}^{-2\pi\iu\nu\tau(\bm{\theta}_{\text{MW},m},\bm{\theta}_{\text{host},l})}
    \label{Eq:IRF_freq}
\end{equation}
or, in more general notation of a field $f$ of complex amplitudes over all paths:
\begin{subequations}
    \begin{align}
        R(t) &= \int\, f(\bm{\theta}_{\text{MW}},\bm{\theta}_{\text{host}})\, \delta\left( t - \tau(\bm{\theta}_{\text{MW}},\bm{\theta}_{\text{host}}) \right)  \der^2 \theta_{\text{MW}}\,\der^2 \theta_{\text{host}}\, ,
    \label{Eq:IRF_time_field} \\
        \tilde{R}(\nu) &= \int\, f(\bm{\theta}_{\text{MW}},\bm{\theta}_{\text{host}})\, \text{e}^{-2\pi\iu\nu\tau(\bm{\theta}_{\text{MW}},\bm{\theta}_{\text{host}})}  \der^2 \theta_{\text{MW}}\,\der^2 \theta_{\text{host}}\, .
    \label{Eq:IRF_freq_field}
    \end{align}
\end{subequations}

Not only are numerical convolutions much faster in Fourier space but the receiver is only sensitive to a limited band. Reflecting this fact, computations are performed at baseband -- as are observations -- such that
\begin{equation}
\begin{split}
    (E_{\text{int}} * R)(t) &= \int_{-\infty}^{\infty}  \, \tilde{R}(\nu) \tilde{E}_{\text{int,true band}}(\nu) \,\text{e}^{2\pi\iu\nu t}\der \nu \\
    &= \int_{-\Delta\nu/2}^{\Delta\nu/2} \, \tilde{R}(\nu+\nu_0) \tilde{E}_{\text{int,baseband}}(\nu) \,\text{e}^{2\pi\iu\nu t} \der \nu 
\end{split}
\end{equation}

where {$\nu_0$ is the central frequency of the observed band and $\Delta\nu$ is the bandwidth. This approach reduces the numerical resolution in time to the Nyquist sampling rate, while $E_{\text{int}}(t)$ is a complex quantity now.

The complex amplitude field $f$ needs to be modeled. Due to numerical limitations, as well as the success of the model of frozen discrete images, we draw a limited number of random paths with nonzero amplitude from a uniform distribution. We consider the strong scattering regime where there is no dominant central beam. A reasonable model for $f$ is a random and isotropic distribution whose mean magnitude is a Gaussian of width $\theta_L$ and independent amplitudes on the two screens:

\begin{equation}
\begin{split}
    f(\bm{\theta}_{\text{MW}},\bm{\theta}_{\text{host}}) &= f_{\text{MW}}(\bm{\theta}_{\text{MW}}) \times f_{\text{host}}(\bm{\theta}_{\text{host}}) \, , \\
    \left\langle \vert f_{\text{MW}}(\bm{\theta}_{\text{MW}}) \vert^2 \right\rangle &\propto \exp\left( -\frac{\bm{\theta}_\text{MW}^2}{ \theta_{L,\text{MW}}^2} \right) \, ,\\
    \left\langle \vert f_{\text{host}}(\bm{\theta}_{\text{host}}) \vert^2 \right\rangle &\propto \exp\left( -\frac{\bm{\theta}_\text{host}^2}{\theta_{L,\text{host}}^2} \right) \, .
\end{split}
    \label{Eq:GaussianScatteringDisk}
\end{equation}

The intrinisically point-like source is distorted into a scattering disk of scale $\theta_L$. Its size depends on frequency approximately following
\begin{equation}
    \theta_L \propto \nu^{-2} \, .
    \label{Eq:nu-2scaling}
\end{equation}

Although different from the $\nu^{-2.2}$ scaling expected for Kolmogorov turbulence, this scaling is compatible with observations and has been derived for models of turbulent media \citep{1977ARA&A..15..479R}. Within the stationary phase approximation \citep{walker2004interpretation} -- which leads to discrete images -- the requirement of stationary phase also leads to scattered images being confined to a region whose size evolves according to $\nu^{-2}$. As discussed, this evolution is not applied here for the frequency channels of one data set but only to different data sets of different central frequencies.

The intrinsic signal $E_\text{int}$ of the source can have a complicated shape. To simulate the impact of scattering, two highly idealized models are used. The first possibility is to neglect its extent by starting from a Dirac delta pulse. This is useful to fully separate the effect of scattering. The second model applied here is a complex random field with a Gaussian distribution, modulated with a Gaussian pulse shape. Its smooth and symmetrical shape still allows for a clear distinction of scintillation effects from intrinsic shape while it results in additional contributions and realistic limitations in analysis steps. 

\subsection{Resolution power of screens}
\label{subsec: RP}

As explained in \cref{Sec:Intro}, scintillation is expected to be quenched if the source is resolved as an extended object.

On Earth, the spatial resolution of radio telescopes is not high enough to resolve the emission regions of pulsars that are constrained to $\lesssim 500$\,km \citep{1997A&A...322..846K} and this has also not been accomplished for FRBs yet. However, a plasma screen can function similarly to very long baseline interferometry (VLBI), with each image point on the screen acting as an individual radio telescope, giving the screen an effective aperture $L$ that can be of the order of astronomical units. The smallest resolved features are then of angular size

\begin{equation}
    \theta_{\text{res}} = \frac{\lambda}{L} \, .
\end{equation}

The length $L$ cannot be uniquely defined in our model because of the Gaussian distribution of images. To get a reasonable parameter of the screen resolution, we will define it as the $2\sigma$ width of complex amplitude distribution $f$ defined in \cref{Eq:GaussianScatteringDisk}. Then, the screen in the MW has a length of
\begin{equation}
    L_{\text{MW}} = 4 \,\theta_{L,\text{MW}} D_{\text{MW}} \, .
    \label{Eq:box_size}
\end{equation}
To ensure that we do not accidentally introduce effects of higher resolution by larger separations than this length, we limited simulations in this work to a box of size $L\times L$. Furthermore, we enforced a vanishing deviation from the mean squared modulus given in \cref{Eq:GaussianScatteringDisk} by fixing
\begin{equation}
    f(\bm{\theta}_{\text{MW}},\bm{\theta}_{\text{host}}) = N \,\exp\left( -\frac{\bm{\theta}_\text{MW}^2}{ 2\theta_{L,\text{MW}}^2} \right) \, \exp\left( -\frac{\bm{\theta}_\text{host}^2}{2\theta_{L,\text{host}}^2} \right) \, . \label{Eq:sim_image_amplitudes}
\end{equation}

Complex amplitudes along the screen were assumed to be complex Gaussian random variables above. In practice, we found that using \cref{Eq:sim_image_amplitudes} leads to an equivalent result even though it is neither complex nor random. The uniform random placement of images already creates complex random deviations between nearby images. Each small region in the integral given in \cref{Eq:IRF_freq_field} consists of a sum of complex contributions of images that follows an Irwin-Hall distribution which becomes a Gaussian distribution for a large number of images. 

We introduce a parameter called Resolution Power (RP) for a two-plane system to quantify the effects of resolving screens. RP is defined as the ratio of the angular size of one screen as seen from a second screen ($\Theta_{\text{size}}$) to the angular resolution achieved by the second screen ($\theta_{\text{res}}$) 

\begin{equation}
    \text{RP} = \frac{\Theta_{\text{size}}}{\theta_{\text{res}}} = \frac{L_\text{MW}L_\text{host}}{\lambda D_\text{MW,host}} \, .
    \label{eq:RP_def1}
\end{equation}

We differentiate between the different degrees of resolution:
\begin{equation}
    \begin{split}
    \text{if } \text{RP} &\gg 1, \text{ the screens fully resolve each other;}\\
    \text{if } \text{RP} &= 1, \text{ the screens just resolve each other;}  \\
    \text{if } \text{RP} &\ll 1, \text{ the screens do not resolve each other.}
    \end{split}
    \nonumber
\end{equation}

There are conceptual problems with equating a scattering disk to a physically extended source, because all radiation still originates from the same point-like source. This means each point of the scattering disk has a stable phase relation to all other points, making it a coherent extended source. Therefore, the above reasoning is not valid without further investigation. A more robust way to argue for quenching is to start from the total scattering delay given in \cref{Eq:tau_eff} and the definition of the impulse response function in frequency space in \cref{Eq:IRF_freq_field}. If the mixed term becomes relevant, it effectively means that -- even in the approximation of fixed images -- the contribution from one screen is different for each image on the other screen. As a result, these light paths only add up incoherently and are suppressed. 

Using only the mixed contribution $\tau_\text{mix}$ to the delay and inserting the angular sizes of the screens, we obtain
\begin{equation}
    - 2\pi \nu \tau_\text{mix} = 2\pi \nu \frac{D_\text{eff,MW}}{c} \frac{L_\text{MW}L_\text{host}}{D_\text{MW}D_\text{host}} = 2\pi \,\text{RP}
\end{equation}
which provides another explanation for our choice of RP and sets the argument of resolving screens on firm ground.

\subsubsection{Evolution of RP with redshift and frequency}

\label{sec:Resolving Screens}

In a two-screen scenario, considering the physical size of the screens \citep[1-100s of astronomical units for typical Galactic distances,][]{Brisken_2010,Lofarcensus12022A&A...663A.116W}, the chance of one screen just resolving the other is non-zero up to a redshift of 4 (using Eq. \ref{eq:RP_def2}). The effect is more relevant at lower redshift (\(z < 0.3\)) as illustrated in \cref{fig:RP-z}, which plots the resolution power as a function of redshift for a set of screen sizes (\(L_{\text{MW}}, L_{\text{host}}\)). For a localized FRB, this can be calculated using a cosmological model. The figure suggests that screens resolving each other is more likely for low redshift FRBs. Given that most successful localizations of FRBs are below $z=0.3$, we focus on this parameter space in this work. 

Another important feature to notice is the frequency evolution of RP. Due to the width of the screens scaling like $\theta_L\propto\nu^{-2}$,  the resolution power as defined in \cref{Eq:box_size,eq:RP_def1} scales as

\begin{equation}
    \text{RP} \propto \nu^{-3} \, .
    \label{freq_effect1}
\end{equation}

This means that if complete modulation of the pulse by two screens is observed at a specific central frequency, the scattering screen sizes increase as the observation frequency decreases, and the screens would start appearing extended to each other.

\begin{figure}
    \centering
    \includegraphics[width=1\linewidth]{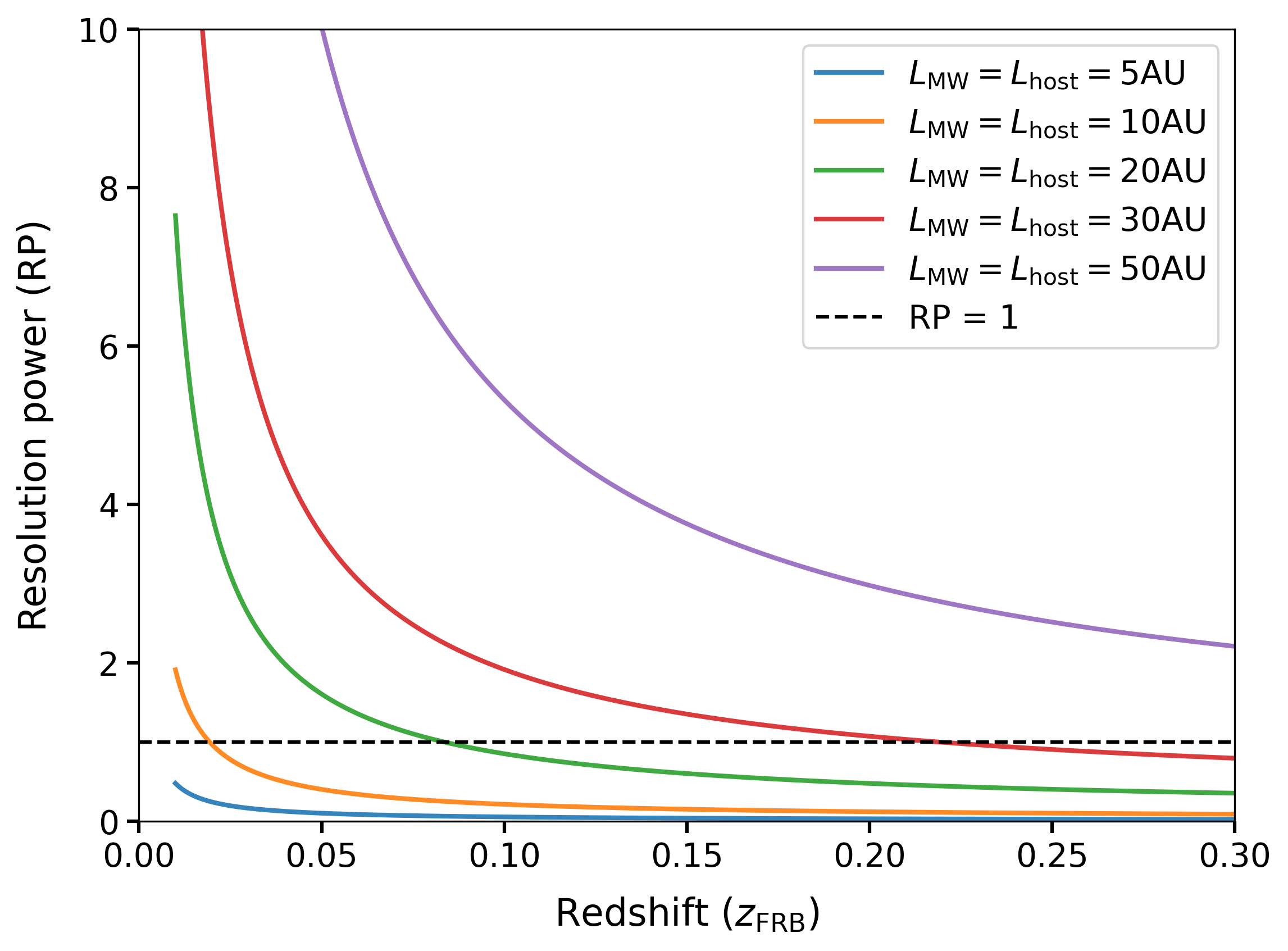}
    \caption{The plot shows the evolution of RP with redshift, where the redshift corresponds to the separation between two screens. The angular diameter distance to the host galaxy (of a given redshift) is calculated assuming a flat $\Lambda$CDM model of cosmology. The screens are fixed to ISMs of the respective galaxies so that we could use the same scattering disk sizes for readability. For each curve, the RP is calculated over a redshift range using equation \ref{eq:RP_def1}. The dashed horizontal line indicates the RP = 1 curve, above which the screens are considered to be resolving each other. One should remember, the resolution power is independent of the screen locations within the galaxy for given screen sizes. A host screen of 50 AU in size, located 1 pc from the FRB, produces a 3 ms scattering delay.}
    \label{fig:RP-z}
\end{figure}

\section{Observables}
\label{Sec: Observables}

In \cref{Sec:CosmoScatteringTheory,Sec:Models}, we described our model which will be used to simulate the electric field. However, the data is usually analysed after being reduced to an intensity that is channelized into a function of time and frequency. For a suitable choice of channel widths, such an array of data is sufficient for the measurement of the observables that will be described in this section. 

The observed intensity as a function of time or frequency is
\begin{equation}
    I(t) \propto \left\vert E(t) \right\vert^2 ~~ \text{or} ~~ \tilde{I}(\nu) \propto \left\vert \tilde{E}(\nu) \right\vert^2 \, .
\end{equation}

This illustrates the fact that the same effect can be observed both in scattering and in scintillation depending on the choice of channelization. In the case of a pulse shaped like a delta function the intensity is just $I\propto |R|^2$, $R$ being the scatter response function. For simplicity, we will start from this case. For scintillation, only relative intensities matter. Hence, we define constants such that
\begin{equation}
    I = |R|^2 \, .
\end{equation}
In the following, we will also simplify notations by denoting Fourier conjugates with the same symbol, e.g.~ $\tilde{I}(\nu)\mapsto I(\nu)$. Scattered images are considered to be randomly placed in a box whose size was defined in \cref{Eq:box_size}. Hence, they represent a sample of a random distribution obeying \cref{Eq:GaussianScatteringDisk}, such that observables are random variables of which only the expectation values will be derived. Quantifying the expected deviations from these values is left for future work.

\subsection{Scattering time}

Phenomenologically, temporal scattering is referred to as the broadening of a burst in time. Usually, it is observed as a scattering tail whose shape is close to an exponential. The expectation value of the intensity measured over time for the model of a delta function shaped burst is
\begin{equation}
    \left\langle I(t) \right\rangle = \left\langle \vert R(t) \vert^2 \right\rangle \, .
    \label{Eq:IntensityDelta}
\end{equation}
The points on the scattering screens are uncorrelated to each other:
\begin{equation}
\begin{split}
    &\left\langle f(\bm{\theta}_{\text{MW}},\bm{\theta}_{\text{host}}) \conj{f}(\bm{\theta}_{\text{MW}}',\bm{\theta}_{\text{host}}') \right\rangle = \\ &\left\langle \left\vert f(\bm{\theta}_{\text{MW}},\bm{\theta}_{\text{host}}) \right\vert^2 \right\rangle \delta^{(2)}\left( \bm{\theta}_{\text{MW}}-\bm{\theta}_{\text{MW}}' \right)\delta^{(2)}\left( \bm{\theta}_{\text{host}}-\bm{\theta}_{\text{host}}' \right)\, .
\end{split}
\label{Eq:ScatteringDiskCorr}
\end{equation}
Inserting \cref{Eq:ScatteringDiskCorr,Eq:IRF_time_field} into \cref{Eq:IntensityDelta}, we obtain
\begin{equation}
\left\langle I(t) \right\rangle = \int\, \left\langle \left\vert f(\bm{\theta}_{\text{MW}},\bm{\theta}_{\text{host}}) \right\vert^2\right\rangle\, \delta^2\left( t - \tau(\bm{\theta}_{\text{MW}},\bm{\theta}_{\text{host}}) \right) \der^2 \theta_{\text{MW}}\,\der^2 \theta_{\text{host}} \, .
\end{equation}
The squared modulus of the screens' amplitudes $f$ is given in \cref{Eq:GaussianScatteringDisk} and the delay $\tau$ is given in \cref{Eq:tau_eff}. Thus, we obtain
\begin{equation}
\begin{split}
    \left\langle I(t) \right\rangle &\propto \int\der^2 \theta_{\text{MW}}\,\int\der^2 \theta_{\text{host}}\, \exp\left( -\frac{\bm{\theta}_\text{MW}^2}{ \theta_{L,\text{MW}}^2} \right)\exp\left( -\frac{\bm{\theta}_\text{host}^2}{\theta_{L,\text{host}}^2} \right) \times\\
    & \delta^2\left( t - \left[ \frac{D_{\text{eff,MW}}}{2c} \bm{\theta}_{\text{MW}}^2 - \frac{D_{\text{eff,MW}}}{c}\bm{\theta}_{\text{MW}}\cdot\bm{\theta}_{\text{host}} + \frac{D_{\text{eff,host}}}{2c} \bm{\theta}_{\text{host}}^2 \right] \right) \, .
\end{split}
\end{equation}
Often, contributions to the delay dominantly come from the host screen \citep{masui2015dense,sammons2023two}. Then, the equation above factorizes into two separate integrals over each screen because $\bm{\theta}_{\text{MW}}$ is no longer relevant in the delta distribution and the image amplitudes on each screen are independent of each other:
\begin{equation}
    \begin{split}
        \left\langle I(t) \right\rangle &\propto \int\, \exp\left( -\frac{\bm{\theta}_\text{host}^2}{\theta_{L,\text{host}}^2} \right)\, \delta^2\left( t - \frac{D_{\text{eff,host}}}{2c} \bm{\theta}_{\text{host}}^2 \right) \der^2 \theta_{\text{host}} \\
        &\propto \exp\left( -\frac{2c t}{D_{\text{eff,host}} \theta_{L,\text{host}}^2} \right)
        \label{Eq:der_scattering_tail}
    \end{split}
\end{equation}

which indeed is an exponential.
The scattering time $\tau_\text{s}$ is defined such that
\begin{equation}
    \left\langle I(t) \right\rangle \propto \exp\left( -\frac{t}{\tau_\text{s}} \right) \, . \label{Eq:def_scattering_tail}
\end{equation}
Thus, equating \cref{Eq:der_scattering_tail} and \cref{Eq:def_scattering_tail} the scattering time while neglecting the MW screen is
\begin{equation}
    \tau_\text{s,host} = \frac{D_{\text{eff,host}} }{2c}\theta_{L,\text{host}}^2 \, .
    \label{Eq:tau_s_host}
\end{equation}

Scattering tails can only be measured when their time scale is comparable to or larger than the intrinsic burst duration as long as the intrinsic burst structure remains unknown. Scattering of smaller delays becomes visible as scintillation over frequency if the additional condition is fulfilled that the scattered rays are coherent.

\subsection{Scintillation bandwidth}
\label{Sec:nu_scint}

The autocorrelation function (ACF) of scintillation as a function of frequency lags is often measured as Lorentzian distribution around zero -- although sometimes a Gaussian fit is used -- whose width provides a characteristic scale of scintillation which is called the scintillation bandwidth or decorrelation bandwidth. \citet{1998ApJ...505..928G} derived this functional form for a single thin screen. In absence of an analytic solution, Lorentzians have also been used to fit the individual components in a system of two screens when the impact of one of them was considered negligible. In a recent observational study, \citet{2025Natur.637...48N} analyzed the total spectral ACF of a burst encountering two screens as a sum of individual Lorentzian components. Here, we present a derivation of analytical solutions for the ACF that is not only valid when one of the screens is negligible but also for two scattering screens that do not resolve each other.

The ACF normalized by the mean is defined as
\begin{equation}
    \text{ACF}(\nu_1,\nu_2) = \left\langle\frac{ I(\nu_1)-\langle I(\nu_1) \rangle}{\langle I(\nu_1) \rangle} \times \frac{ I(\nu_2)-\langle I(\nu_2) \rangle}{\langle I(\nu_2) \rangle} \right\rangle \, .
    \label{Eq:def_ACF}
\end{equation}
Now we use the complex Isserlis theorem \citep{1974sats.book.....K} and $I\propto |R|^2$ to obtain
\begin{equation}
\begin{split}
    \langle I(\nu_1) I(\nu_2) \rangle &=
\langle R(\nu_1) \conj{R}(\nu_1) R(\nu_2) \conj{R}(\nu_2) \rangle
  \\ &=
  \langle R_1 \conj{R_1}\rangle\langle R_2 \conj{R_2} \rangle + 
  \langle R_1 \conj{R_2}\rangle\langle  R_2 \conj{R_1} \rangle
  \\ &=
  \langle I(\nu_1)\rangle \langle I(\nu_2) \rangle + \left\vert\langle R_1 \conj{R_2}\rangle\right\vert^2 \, .
\end{split}
\label{Eq:Isserlis}
\end{equation}
where for brevity the shortcut notation of $R(\nu_1)=R_1$ and $R(\nu_2)=R_2$ was used. We will also abbreviate $\delta\nu = \nu_1-\nu_2$.
Inserting \cref{Eq:IRF_freq_field} yields
\begin{equation}
    \langle R_1 \conj{R_2}\rangle = \int\, \left\langle \left\vert f(\bm{\theta}_{\text{MW}},\bm{\theta}_{\text{host}}) \right\vert^2 \right\rangle\, \text{e}^{-2\pi\iu\delta\nu\tau(\bm{\theta}_{\text{MW}},\bm{\theta}_{\text{host}})} \der^2 \theta_{\text{MW}}\,\der^2 \theta_{\text{host}}\, .
    \label{Eq:Response_Corr}
\end{equation}

First, we look at the case where the contribution of one of the screens dominates in the delay. Then we define for this dominating angle $\bm{\theta}$
\begin{equation}
    \tau = \frac{D_\text{eff}}{2c}\theta^2 \, .
\end{equation}
At this point the scintillation bandwidth $\nu_\text{scint}$ -- or $\nu_\text{s}$ for brevity -- is introduced as
\begin{equation}
    \nu_\text{s} = \frac{c}{\pi D_\text{eff} \theta_L^2} \, .
    \label{Eq:nu_s_MW}
\end{equation}
Comparison with \cref{Eq:tau_s_host} shows that scattering time and scintillation bandwidth relate to each other like
\begin{equation}
    \nu_\text{s} = \frac{1}{2\pi\tau_\text{s}}  \, .
    \label{Eq:delay_to_bandwidth}
\end{equation}
This fact has been used to check if measured values correspond to the same screen or indicate two different screens \citep{masui2015dense,sammons2023two}.

In \cref{Eq:Response_Corr}, neglecting one screen again leads to the integral factorizing where
the scattering angle whose screen has been neglected can be separated into an integral yielding a constant. In polar coordinates only the integration along the radius remains which can be expressed in terms of the delay $\tau$:

\begin{equation}
    \langle R_1 \conj{R_2}\rangle \propto \int_0^\infty\, \exp\left( -2\pi\nu_\text{s}\tau-2\pi\iu\delta\nu\,\tau\right)\,\der\tau \, .
\end{equation}
The solution of this Fourier integral is 

\begin{equation}
    \langle R_1 \conj{R_2}\rangle \propto \frac{1}{\nu_\text{s}+\iu \delta\nu} \, .
    \label{Eq:IRF_crosscorr}
\end{equation}
To compute the ACF as defined in \cref{Eq:def_ACF}, we need to also compute the expectation value of the intensity. This is equivalent to setting $\nu_1=\nu_2$ in \cref{Eq:IRF_crosscorr}:
\begin{equation}
    \langle I \rangle \propto \frac{1}{\nu_\text{s}}
    \label{Eq:I_expect}
\end{equation}
where the constant of proportionality is the same as in \cref{Eq:IRF_crosscorr}. Finally, combining \cref{Eq:def_ACF,Eq:Isserlis,Eq:IRF_crosscorr,Eq:I_expect} we obtain the ACF for this case:
\begin{equation}
    \text{ACF}(\nu_1,\nu_2) = \frac{\left\vert\langle R_1 \conj{R_2}\rangle\right\vert^2}{\langle I \rangle^2} = \frac{1}{1+\left( \delta\nu/\nu_\text{s} \right)^2} \, .
    \label{Eq:ACF_Lorentzian}
\end{equation}

If both screens are considered but do not resolve each other, i.e.~the cross term containing angles on both screens vanishes, the response function can be split into two statistically independent factors:
\begin{equation}
    R(\nu) = R_\text{MW}(\nu) \, \times \, R_\text{host}(\nu) \, .
\end{equation}
In fact, we can generalize to an arbitrary number of screens labeled by $n$, which leads to a general expression for the ACF analogous to the derivation for a single screen:
\begin{equation}
    \begin{split}
        1 + \text{ACF} &= \frac{\langle I(\nu_1) I(\nu_2) \rangle}{\langle I \rangle^2} \\
        &= \prod_n \frac{\langle R_n(\nu_1) \conj{R_n}(\nu_1) R_n(\nu_2) \conj{R_n}(\nu_2) \rangle}{\langle I_n \rangle^2} \\
        &= \prod_n \left( 1+\text{ACF}_n \right) \, .
    \end{split}
    \label{Eq:Nscreen_ACF}
\end{equation}
The implications of this result for two screens are illustrated in \cref{fig:theo_TwoScreenACF} and result in a combination of additive and multiplicative components:
\begin{equation}
    \text{ACF} = \text{ACF}_\text{MW}\times\text{ACF}_\text{host} + \text{ACF}_\text{MW} + \text{ACF}_\text{host}
    \label{Eq:TwoScreenACF}
\end{equation}
The multiplicative component can lead to ACF shapes deviating from Lorentzian shapes if the scales of both contributions are similar. Until now, studies like \citet{2025Natur.637...48N} formed models by adding Lorentzian components instead.

\begin{figure}
    \centering
    \includegraphics[width=1\columnwidth]{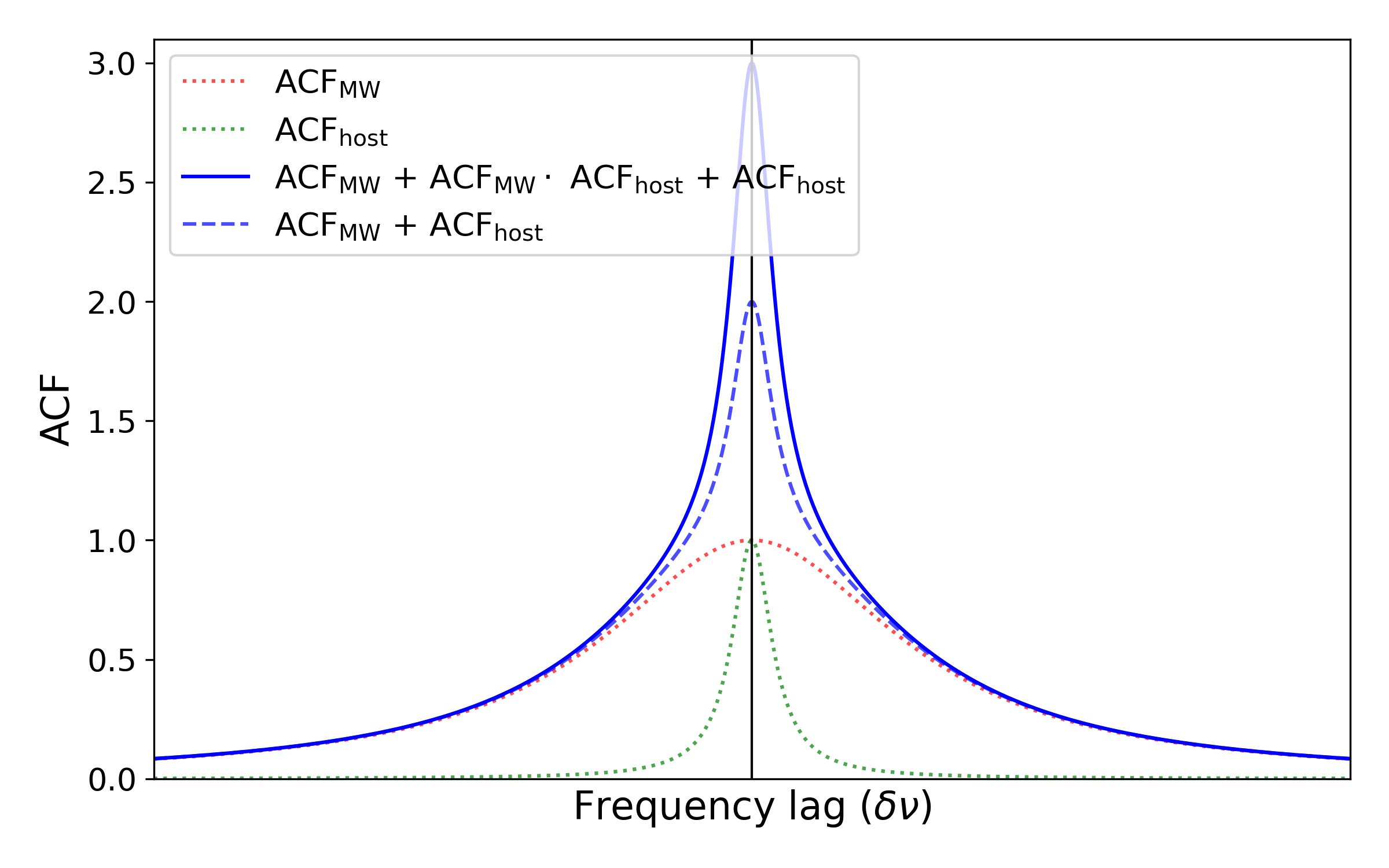}
    \caption{Illustration of the different contributions to the ACF as given in \cref{Eq:TwoScreenACF} in the case of two screens that do not resolve each other. Lorentzians expected from only the MW screen (red) or only the host screen (green) do not add up (dashed blue) to the correct two-screen distribution (solid blue).}
    \label{fig:theo_TwoScreenACF}
\end{figure}

In practice, the scintillation bandwidth is measured by fitting a single Lorentzian function as given in \cref{Eq:ACF_Lorentzian} even in the presence of two screens. It is obtained as the half width at half maximum (HWHM) of such a fit. As can be seen in \cref{Eq:TwoScreenACF}, this approach is still correct for the terms belonging to the much wider component of two ACF components as long as the center is omitted (see Fig.~\ref{fig:theo_TwoScreenACF}). This central part is defined by the narrower ACF component being nonzero, for which another fit function has to be used:

\begin{equation}
    \text{ACF} \simeq \frac{2}{1+\left( \delta\nu/\nu_\text{s} \right)^2} + 1 \, .
\end{equation}
To obtain general and robust fits, we left the prefactor and additive constant as free parameters in our numerical analysis.

Measuring the scintillation bandwidths of both screens simultaneously can require very narrow channels in order to resolve the narrow scintles of the second screen. If not resolved, the smaller $\nu_s$ can be reduced to a peak at a lag of zero. This is problematic for two reasons: First, this position is contaminated by noise that is not correlated over frequency. Second, the width of the intrinsic burst shape also produces a very small decorrelation bandwidth. Due to the inverse relation between scattering delay and scintillation bandwidth given in \cref{Eq:delay_to_bandwidth}, a small $\nu_s$ corresponds to a large $\tau_s$ that might be comparable to the time scale of the burst in which case both components overlap in the ACF.

The impact of an intrinsic burst profile that is not a delta function can also be evaluated using \cref{Eq:Nscreen_ACF}. In the frequency domain, the measured electric field as defined in \cref{Eq:def_E_measured} becomes $E(\nu) = E_{\text{int}}(\nu) R(\nu)$ such that the Fourier transform of the intrinsic profile can be treated like the response function of an additional screen. The same is true for instrumental bandpass functions, in case they have not been divided out before the analysis. Any contribution with a very wide ACF will modify the amplitude of the narrower ACFs because of the additional product terms.

\subsection{Modulation index}
\label{sec:mod_index}

The modulation index is defined as the relative standard deviation of the intensity:

\begin{equation}
    m = \frac{\sigma_I}{\expv{I}} \, ,
    \label{eq:mod_1}
\end{equation}
where $\sigma_I$ is the standard deviation and $\expv{I}$ is the mean of the intensity. Within the context of this paper, we consider the modulation index as being measured along frequency and not along time. The modulation index is an easy to measure probe of the strength of scintillation \citep{1990ARA&A..28..561R}, typically ranging from 0 -- no scintillation -- to 1 -- strong scintillation. Here, we show that a modulation index above 1 is a strong indicator for multiple screens. In practice, a modulation index measured from data will also include additional contributions from instrumental effects such as radiometer noise or baseline variations, as well as radio frequency interference. Any additional intrinsic spectral structure, for example, due to band-limited emission often observed in FRBs will add additional contributions to the measured $m$. 
 
The modulation index can also be obtained from the ACF. Comparing to the definition of the ACF in \cref{Eq:def_ACF}, we obtain the relation
\begin{equation}
    m^2 = \text{ACF}(\delta\nu=0) \, .
\end{equation}
Hence, the modulation index can be alternatively obtained from the peak value of the ACF, or, as shown in \cref{fig:theo_TwoScreenACF}, from the peak of the fitted Lorentzian, which is less biased by instrumental or intrinsic contributions. From the results given in \cref{Eq:ACF_Lorentzian,Eq:TwoScreenACF} follows that $m=1$ for a single screen and $m=\sqrt{3}$ for two screens that do not resolve each other. For $N$ screens, \cref{Eq:Nscreen_ACF} leads to the general formula
\begin{equation}
    m^2 = 2^N-1 \, .
    \label{eq:m_I_nscreen}
\end{equation}

The fully resolved case of two-screen scintillation is equivalent to the single-screen case leading to $m=1$. This can be understood by considering each path through both screens as independent of each other as a consequence of the non-vanishing mixed term in the delay defined in \cref{Eq:tau_eff}. Effectively, each full path can be treated like as a different image on a single screen, such that the result for the modulation index is the same. The effect on the ACF shall be explored numerically in the following sections.

\cite{2024PNAS..12106783J} derived a formula for the modulation index where it is proportional to the inverse number of images on the screen responsible for the narrow scintillation instead of depending on the number of screens as derived here. They argue that, in the resolved regime (RP >> 1), each image in the host screen is independently modulated by the MW screen, while the narrow scintillation is not observed due to being narrower than the frequency channels. In contrast, we assume a large number of images on each screen and sufficiently fine channels, which explains the different results. In the completely resolved regime and in the limit of a large number of images, $m_{\text{MW}}$ approaches zero, which aligns with our results.

\section{Data simulation and analysis methods}
\label{Sec: FRB_Scintillator}

In this section, we detail the implementation of the theoretical framework outlined in \cref{Sec:CosmoScatteringTheory} within the \texttt{FRB\_scintillator} code. The data simulation using {\tt FRB\_scintillator} has two phases: \romannum{1}) calculating the complex screen impulse response function (IRF) using \cref{Eq:IRF_time} and \romannum{2})  convolving the IRF with the simulated, intrinsic pulse of choice.

The spatial distribution, amplitudes distribution and number of images on a scattering screen are key properties that significantly influence the IRF. An approximation used in our simulations are the use of a finite number of images in a screen. In theory, there could be hundreds of thousands of image points on the screen, but a model with \(\sim10^2\) images agrees with the number of distinct features in observations
of pulsars like PSR B0834+06 \citep{Brisken_2010,2021MNRAS.500.1114S} and hence realistically models a screen's impact on the pulse. As mentioned in \cref{Sec:Models}, the screen is modeled with images randomly distributed in the plane, with a Gaussian intensity distribution centered on the optical axis to characterize the image strength (Fig. \ref{fig:2-Screen_image_dist}). In a two-screen system, with \(N_1\) and \(N_2\) images on each screen respectively, there are \(N_1 \times N_2\) propagation paths, each contributing a corresponding delay to the simulation of the IRF (Eq. \ref{Eq:IRF_time}). An example of a simulated IRF is shown in \cref{fig:ResponseFunction}, where each vertical line represents a distinct propagation path.

The screen size and its location between the source and observer are other important parameters that governs the IRF. From the illustration of an FRB encountering a two-screen system shown in \cref{fig:2screen_illustration}, we have four planes of interest: the Observer plane (Obs), the MW screen plane, the Host galaxy screen plane, and the FRB source plane (FRB). The associated free parameters used in simulating the IRF are the redshifts of each plane \(z_{\text{src}}\ = z_{\text{host}}\) and \(z_{\text{MW}} = z_{\text{Obs}} = 0\), the distance between the host screen and the source \(D_{\text{sh}}\) and the distance between the MW screen and the source \(D_{\text{MW}}\). The simulation employs a fixed screen formulation, where the image locations remain constant throughout the frequency band of simulation. Consequently, the additional free parameters describing the screens are their sizes, \(L_{\text{host}}\) and \(L_{\text{MW}}\), which are defined at a central frequency \(\nu_0\). Furthermore, this approximation holds over a bandwidth \(\Delta\nu\), within which the screen sizes are assumed to remain unchanged. We also ensure that $\Delta\nu$ is large enough to observe more than fifty scintles, so that the finite scintle error (e.g.~Eq. 3 in \citealp{Turner2021ApJ...917...10T}) is less than 10 percent. 

Screens with sizes between one and tens of astronomical units (AUs) are sufficiently large to probe the resolving effects of two-screen systems in Fast Radio Bursts (FRBs) (see Sec.~\ref{sec:Resolving Screens}). Pulsar studies have shown such large screens in the Milky Way ISM, including a 23 AU screen at 1.4 GHz towards \(\text{PSR J1057--5226}\) \citep{kerr2018MNRAS.474.4637K} and a 16 AU screen at 300 MHz towards \(\text{PSR B0834+06}\) \citep{Brisken_2010}. Hence, most of the simulations in this work use a central frequency of 800 MHz, where most FRBs are detected and AU-sized screens can be expected. 

\begin{figure}
    \centering
    \includegraphics[width=\linewidth]{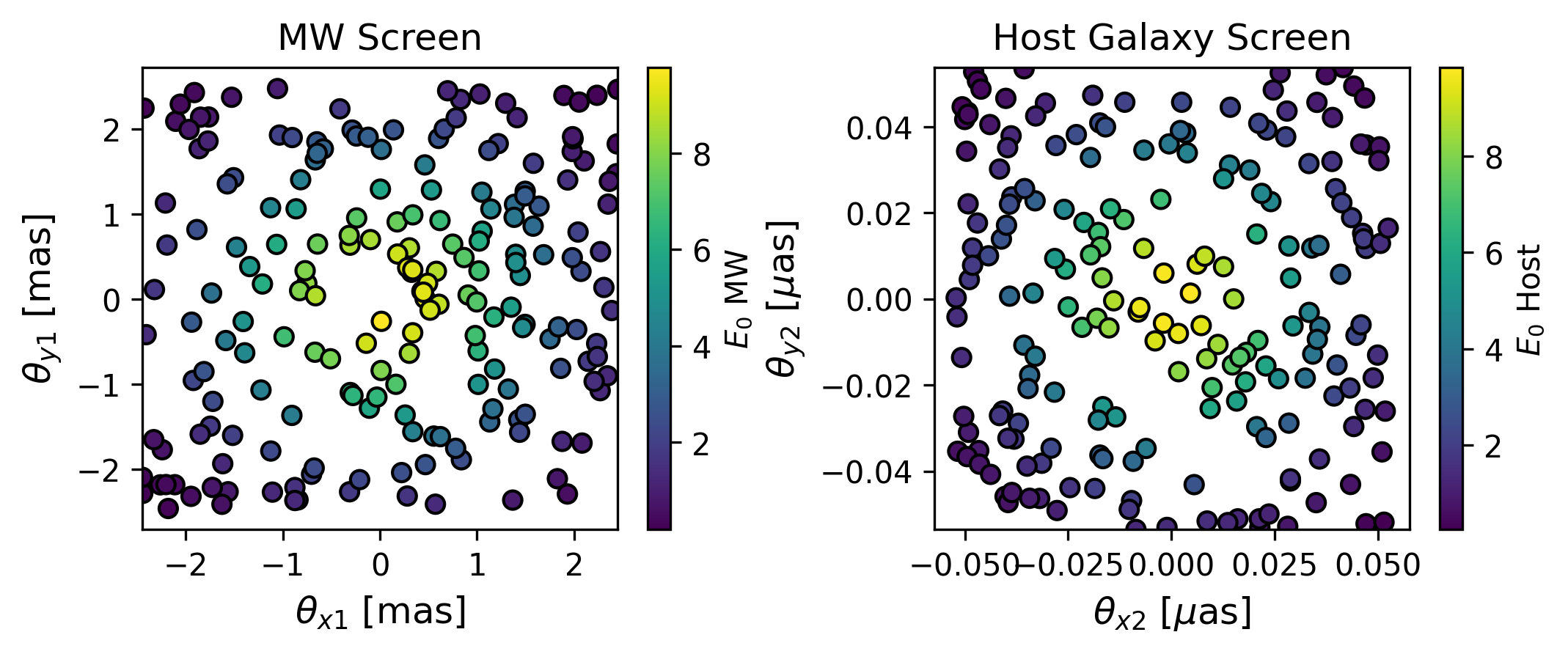}
    \caption{An example MW and host galaxy screen image distribution with color bar showing Gaussian electric field amplitude. Each disk in the plot is an image point with coordinates $\theta_{\text{x}}$ and $\theta_{\text{y}}$. }
    \label{fig:2-Screen_image_dist}
\end{figure}

\begin{figure}
    \centering
    \includegraphics[width=\columnwidth]{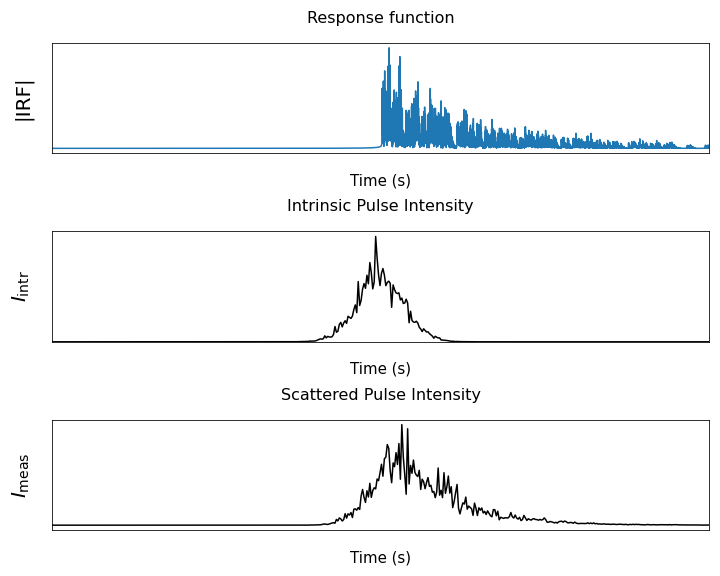}
    \caption{\textbf{Top:} Intensity of the response function as a function of time in seconds. The discrete image distribution manifests as spikes in the response function. The x coordinate of each point corresponds to the delay of a scatter path, and y-coordinate reflects the strength of the pulse traveling through that path. \textbf{Middle:} plot presents the intrinsic pulse intensity profile, while the \textbf{Bottom:} plot illustrates the scattered pulse's intensity profile, characterized by its distinct tail. The last plot was produced by averaging in time to provide a qualitative view of the profiles.}
    \label{fig:ResponseFunction}
\end{figure}

In this study, we generated intrinsic pulses with either Gaussian or Delta-function temporal profiles and flat frequency spectra. We use these pulses to analyze observable effects in the dynamic spectra and the spectral ACF, respectively. Our methods are detailed below.

\begin{enumerate}
    \item \textbf{Gaussian pulse:} This involves examining the visual effects of two screen scattering in "filterbank" format data. To achieve this, the simulated IRF is convolved with a Gaussian intrinsic pulse to generate the time series of a scattered pulse. The resulting time series is then channelized and squared to produce dynamic spectra. Channelization ensures that broader scintillation features are resolved in frequency. In the dynamic spectra, deviations from consistent scintillation patterns across the pulse time profile are analyzed.

    \item \textbf{Delta-function pulse: } For the simplicity in making quantitative measurements, we use a delta function as the intrinsic pulse. Here the time series of the scattered pulse is essentially the IRF.
     
    \begin{enumerate}
        \item First, we apply a Fourier Transform to the simulated time series to obtain the spectrum at highest resolution, and then compute its ACF. We use the  the mean normalized auto-correlation function in order to extract the modulation index from the ACF. Since we use the full spectrum to produce the ACF we call this the \emph{full-spectrum ACF} in this paper.
        \item We fit a Lorentzian function to the ACF to extract the characteristic scintillation bandwidth (\(\nu_{\text{s}}\) or \(\nu_{\text{dc}}\)), which is the half-width at half-maximum (HWHM), and modulation index (\(m\)) of the scintillation scale which is the square root of the peak correlation. The fit function is of the form
        
        \begin{equation}
            f(\delta \nu)= \frac{m^2}{1+ \left(\frac{\delta \nu}{\text{HWHM}}\right)^2} + C
        \end{equation}
        where \(\delta \nu\) is the frequency lag, \(m^2\) is the peak correlation, and \(C\) is an arbitrary constant. The fit is applied by isolating each visually distinct component of the measured ACF using a threshold applied to the correlation coefficient values.
        
        \item Next, we compare the extracted scintillation parameters with the injected values and theoretical expectations. Throughout this paper, the term `injected values' refers to the delays caused by individual screens, which are calculated from the screen parameters using \cref{Eq:delay_formula}). 
        \end{enumerate}
    \item If the dynamic spectra shows a change in the scintillation pattern across the pulse profile, one can quantify this phenomenon in two ways. We divide the pulse dynamic spectrum into spectra as a function of time bins.
    \begin{enumerate}
        \item Then we compute the spectral auto-correlation function for each time bin and measure the \(\nu_{\text{s}}\) of the broad scale scintillation.
        \item In a similar fashion, we also measure the modulation index, \(m\), evolution across the pulse profile. \(m\)(t) is calculated using the peak ACF of the spectra in each time bin \citep{sammons2023two}.
    \end{enumerate}
\end{enumerate}

\section{Results: Two screen simulations}
\label{Sec: Simulation Results}

We use the simulation tool to create data for the three broad categories specified in \cref{subsec: RP}, namely: unresolved, just resolved, and completely resolved screens. The source parameters for each simulation are chosen based on known, localized FRBs with redshifts measured from their host galaxies. We calculate the corresponding angular diameter distance using a flat $\Lambda$CDM model with a Hubble constant of $H_0 \approx 68~\text{km/s/Mpc}$ and a matter density parameter of $\Omega_{m0} = 0.315$ \citep{Planck2020A&A...641A...6P}. These fiducial cosmological parameters are adopted for simplicity, and the precision of the Hubble constant does not significantly impact the distance estimation to the host galaxy. The screen sizes are chosen based on the desired resolution power of the screen system ensuring that the produced delays are of different scales, so that each screen produces a distinct Lorentzian in the full-spectrum ACF. If the delay scales are comparable the ACF would produce a combined single curve making it difficult to extract individual screen parameters. Additionally, we also ensure that the host galaxy screen produces larger delays than the MW screen, following the majority of the observational results so far. Hence, in this work, the MW screen contributes to the broad scintillation, and the host galaxy screen contributes to the narrow scintillation in the ACF. This common assumption, although not always true (e.g., \cite{2025Natur.637...48N}), facilitates a one-to-one comparison with previous FRB two-screen studies in the literature. The free parameters and injected values for each simulation in this section can be found in \cref{sec: Simulation parameters}.

\subsection{Screens that do not resolve each other}
\label{subsec:Unresolved_Screens}

For this case we use the localization of the first repeater FRB 121102A to fix the distance. The FRB was localized to a dwarf galaxy at a redshift of \(z_{\text{src}}\approx 0.192\) \citep{tendulkar2017host}, which is used to fix the source's distance in this simulation. \citet{Ocker2021ApJ...911..102O} located the screen responsible for the scintillation in FRB 121102A within a Milky Way spiral arm at \(\approx 2.3^{+3.2}_{-0.7}\) kpc. Similarly to the MW screen, the host galaxy screen is modeled within the host galaxy's ISM.

\subsubsection{Delta peak as intrinsic pulse}
\label{sec:1.1}

\begin{figure*}
    \centering

        \begin{subfigure}{\linewidth}
        \centering
        \includegraphics[width=\linewidth]{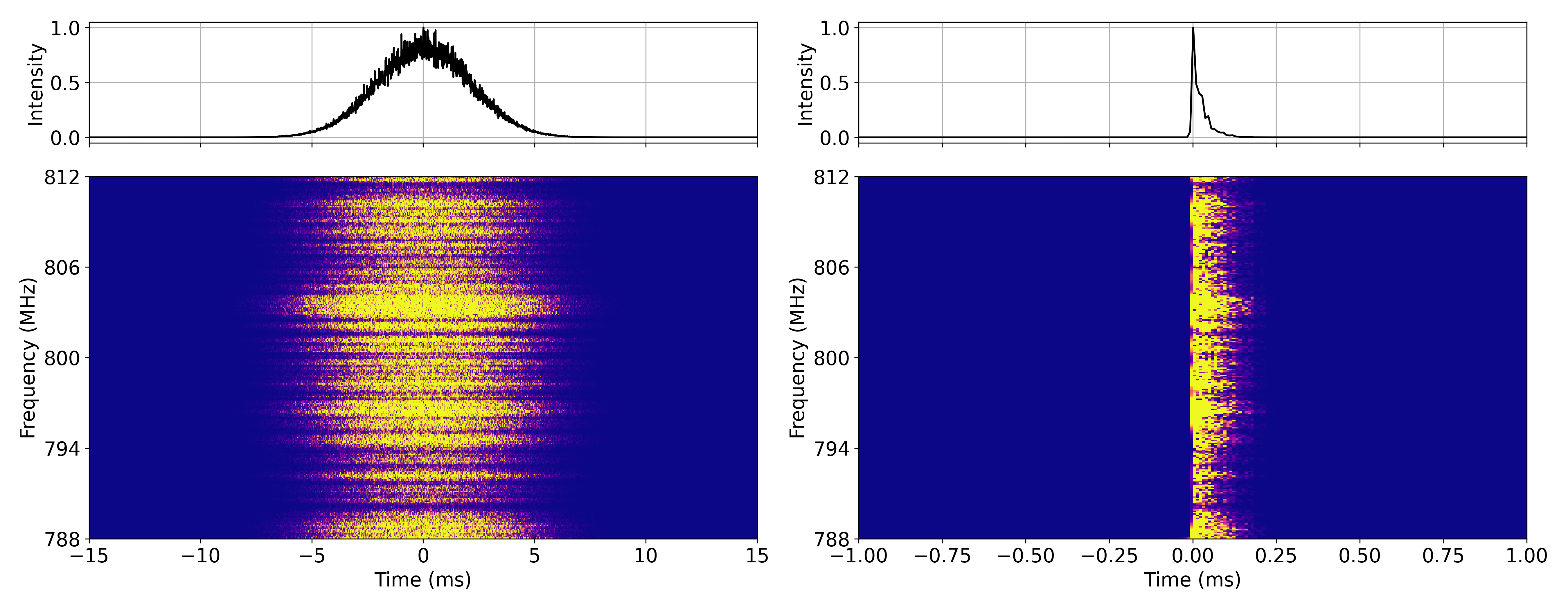}
    \end{subfigure}
    
    \begin{subfigure}{\linewidth}
        \centering
        \includegraphics[width=\linewidth]{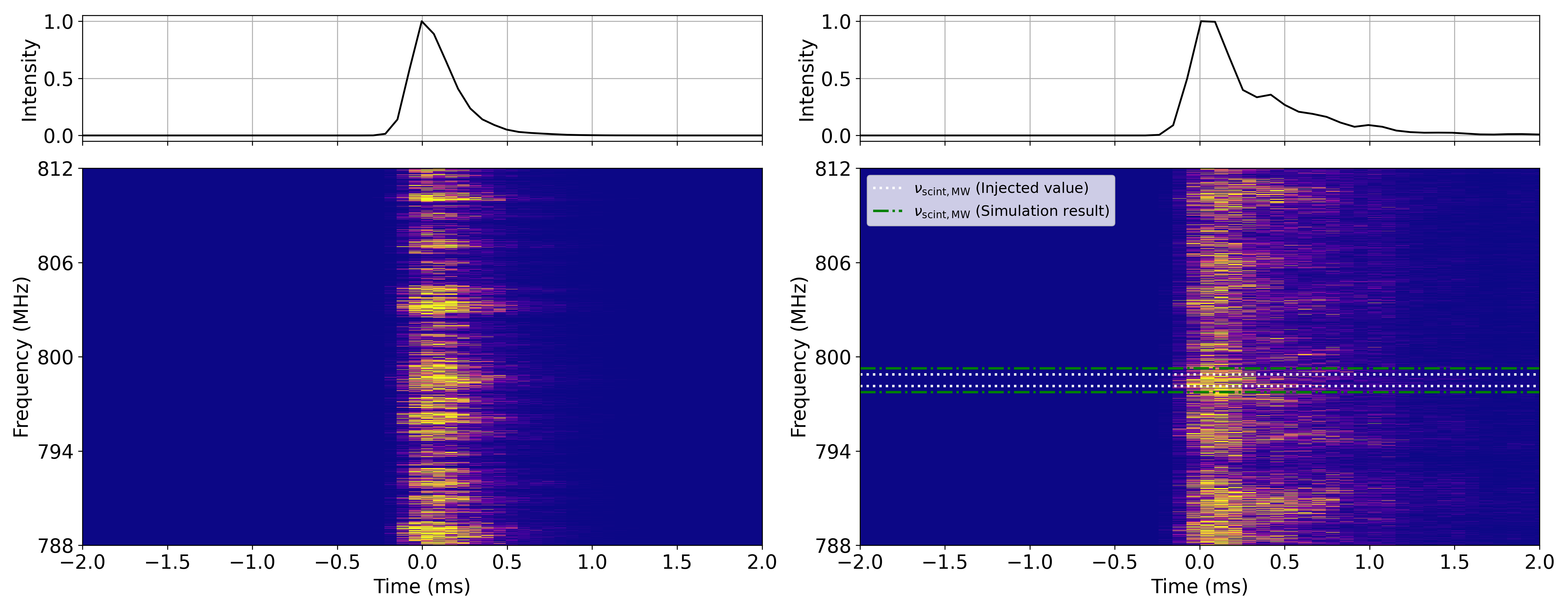}
    \end{subfigure}
    
    \caption{Simulated time profiles and dynamic spectra of a pulse after propagating through unresolved, just-resolved, and completely resolved two-screen systems. \textbf{Top left:} A Gaussian intrinsic pulse after propagating through an unresolved system (RP=0.2). \textbf{Top right:} A delta-function intrinsic pulse after propagating through the same screen system. \textbf{Bottom left:} A Gaussian intrinsic pulse after propagating through a just-resolved system (RP=1). \textbf{Bottom right:} A Gaussian intrinsic pulse after propagating through a completely resolved system (RP=10). In this plot the dotted lines in the dynamic spectra compare the scintillation bandwidths: the injected value from \cref{tab:parameters_two-screen} (white) and the fit value from \cref{fig:Case3 - ACF} (green). Each plot consists of two panels: \textbf{Bottom panel:} The pulse dynamic spectrum, showing channelization resolving broad scales of scintillation in frequency. \textbf{Top panel:} The normalized sum over frequency.} 
    \label{fig:Dynspec_combined}
\end{figure*}

\begin{figure*}
    \centering
    \begin{subfigure}{\linewidth}
        \centering
        \includegraphics[width=\linewidth]{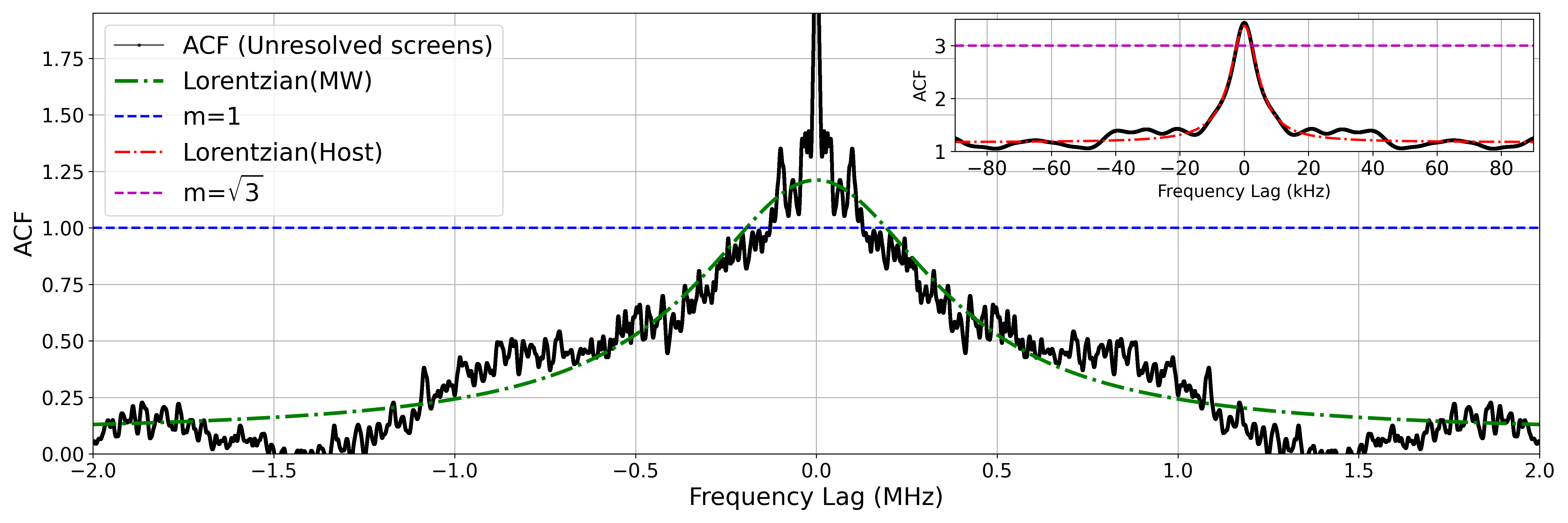}
        \caption{\textbf{Unresolving screens, RP=0.2 }. Discussion in subsection \ref{subsec:Unresolved_Screens}} 
        \label{fig:Case1.1 - ACF}
    \end{subfigure}
    
    \begin{subfigure}{\linewidth}
        \centering
        \includegraphics[width=\linewidth]{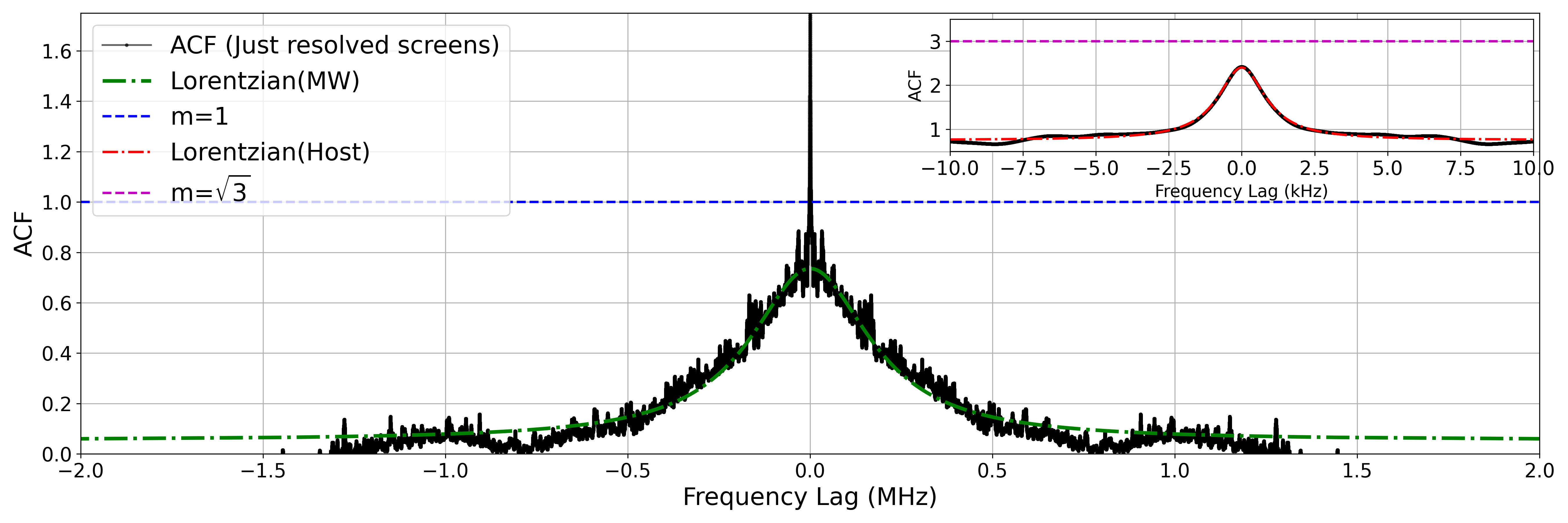}
        \caption{\textbf{Screens that just resolve each other, RP=1 }. Discussion in subsection \ref{subsec:Justresolving}}
        \label{fig:Case2 - ACF}
    \end{subfigure}
    
    \begin{subfigure}{\linewidth}
        \centering
        \includegraphics[width=\linewidth]{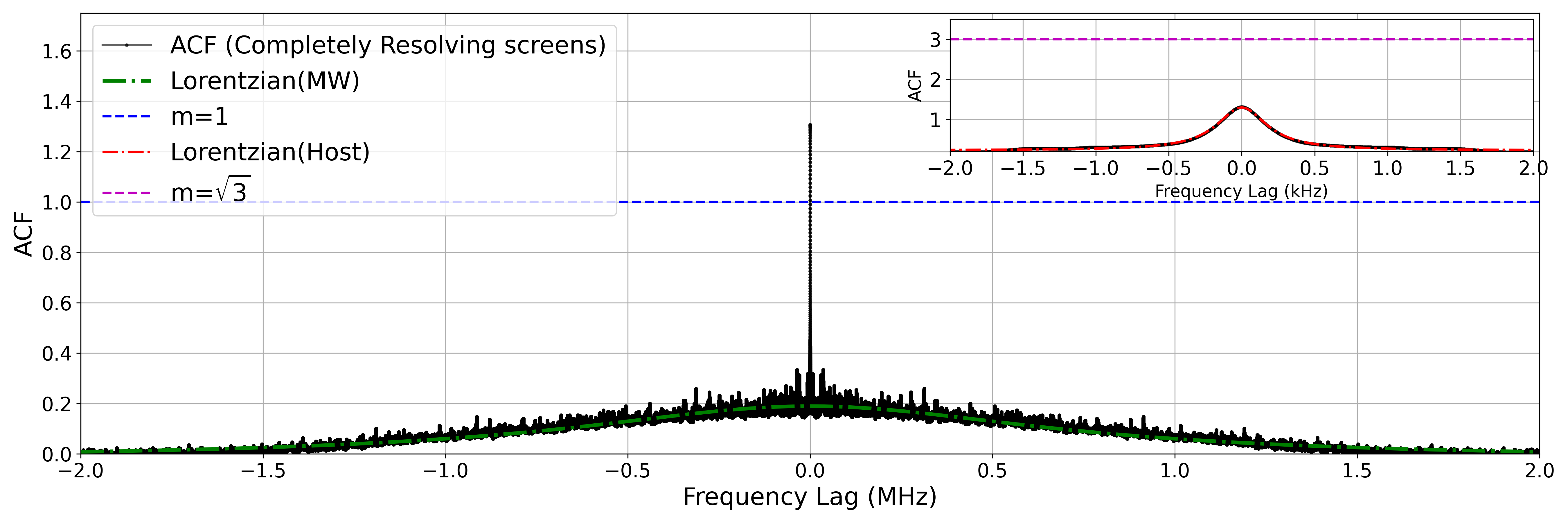}
        \caption{\textbf{Screens that clearly resolve each other, RP=10 }. Discussion in subsection \ref{subsec:resolving_Screens}}
        \label{fig:Case3 - ACF}
    \end{subfigure}
    
    \caption{The plots above show the spectral auto-correlation function (ACF) plotted against the frequency lag. The full-spectrum ACF displays a multi-Lorentzian feature characterizing the presence of multiple screens. \textbf{Main plot:} Broader scintillation produced by the MW screen. \textbf{Inset:} The inset focuses on narrower features in the ACF, highlighting the scintillation produced by the host galaxy screen. Black lines represent simulated data points, while the green and red dash-dotted lines show the Lorentzian function fitted to the ACF of the MW and host scintillation, respectively. The fits are performed by choosing the data points within an ACF range that isolates the respective scale of scintillation. The blue and magenta dashed lines represent the modulation index \(m=1\) and \(m=\sqrt{3}\), corresponding to the complete modulation expected from one screen and combined two screens, respectively. One important feature to note is how the peak ACF of broader scintillation and the combined two screen scintillation drop as we go from the unresolving screen to the resolving screen case. A detailed discussion on this is provided in \cref{sec: Quenching of scintillation}.}
    \label{fig:All_ACFs}
\end{figure*}

A delta function as the intrinsic pulse is the simplest case to study the effect of screens at the observer. Here, the electric field at the observer is essentially the screen's response function. As a first step, we channelize the time series to resolve broad-scale scintillation and generate the dynamic spectra, shown in the top right plot of \cref{fig:Dynspec_combined}. In the dynamic spectrum, we observe a constant scintillation pattern across the pulse time profile, characteristic of a pulse encountering an unresolved two-screen system.

As explained in Sections \ref{Sec: FRB_Scintillator}, to make quantitative measurements, we generate the full spectrum autocorrelation function (ACF) using the same time series, which is shown in \cref{fig:Case1.1 - ACF}. The scintillation bandwidth and modulation index obtained from Lorentzian fits to the individual profiles align well with the injected values ($\nu_{\text{s,MW}}\approx 0.16 \text{MHz}$; $\nu_{\text{s,host}} \approx 250 \text{Hz}$; $m_{\text{MW}}\approx 1$; $m_{\text{two screen}}\approx\sqrt{3}$). It is to be noted that, due to the absence of self-noise, the combined modulation from both screens corresponds to the square root of the peak correlation of the full-spectrum ACF. This result is consistent with our theoretical expectation for the modulation index of a pulse encountering two screens, as derived from the general \textit{N}-screen formula in \cref{eq:m_I_nscreen}. 

The simulations are conducted in the regime where the two scales of scintillations differ by orders of magnitude, producing two distinct curves in the full spectral ACF. We apply thresholding in the correlation to isolate and fit Lorentzians to these two curves, resulting in a wide and a narrow Lorentzian. From equation \cref{Eq:TwoScreenACF}, we know that the total ACF has both additive and multiplicative components. One drawback of our method is that we treat \(\text{ACF}_{\text{host}} + \text{ACF}_{\text{MW}} \times \text{ACF}_{\text{host}}\) as the narrow Lorentzian, a procedure similar to that used in observational studies such as \cite{2025Natur.637...48N}. A host screen modulation index calculation from the peak correlation of the narrow Lorentzian results in an overestimation. Hence, we refrain from discussing \(m_{\text{host}}\) in this work. Since the scintillation bandwidths of the screens differ by orders of magnitude, \(\nu_{\text{s,host}}\) measurements from the fits remain unaffected by this method. A detailed theoretical study on the interplay of narrow scintillation and the product term in the ACF will be presented in an upcoming study.

The errors in the fit are negligible and hence not reported in this section. However, the fit results are approximations due to systematic errors, primarily due to the use of a finite number of images in a screen. The sparse sampling of the screen produces additional bumps over a Lorentzian, resulting in a wiggly Lorentzian profile. These wiggles are similar to noise introduced in the Fourier transform over a sparsely sampled space. Such wiggles are also realistic, as observational results have shown finite number of images in screens \citep{Brisken_2010}. In later sections, where necessary, multiple simulations with varying image distribution seeding have been conducted to account for these systematic errors.

\subsubsection{Gaussian intrinsic pulse and effects of self-noise}
\label{sec:1.0}

\begin{figure*}
    \centering
    \includegraphics[width=\linewidth]{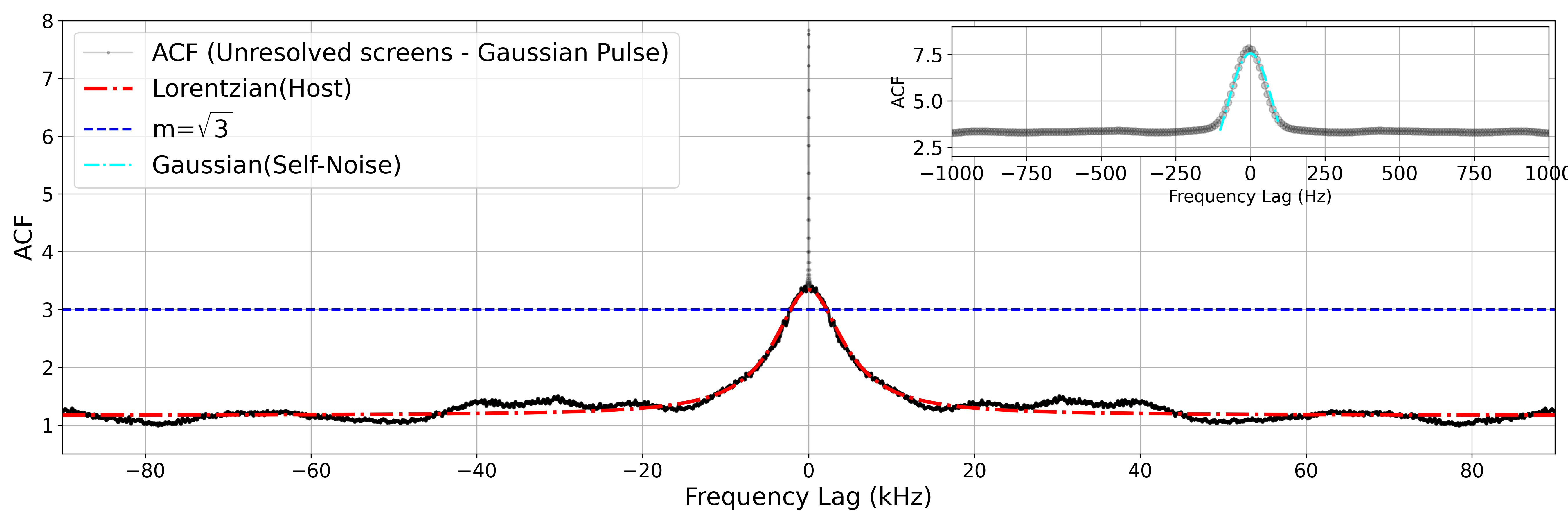}
    \caption{The spectral autocorrelation function (ACF) from the unresolved screen case with an injected Gaussian intrinsic pulse of width 3 ms. The main panel illustrates the host screen scintillation, while the inset highlights the narrower Gaussian self-noise resulting from the injected intrinsic pulse. To see the MW scintillation we refer to \cref{fig:Case1.1 - ACF} as the IRF involved in producing both the plots are the same. The scintillation from the two screens peaks close to 3, while the narrow self-noise peaks higher. This plot emphasizes the importance of distinguishing between scintillation and self-noise when dealing with pulses of finite width. A Gaussian intrinsic pulse results in a Gaussian self-noise curve, but a more complicated intrinsic pulse can produce self-noise of an unknown functional form. Even in such cases, one can estimate the width of the pulse from the self-noise in the spectral ACF.
    }
    \label{fig:Case1_Gaussian_ACF}
\end{figure*}

In reality, a burst will have a finite temporal width, which introduces self-noise in the spectral ACF. For the simplest case of a Gaussian intrinsic pulse with a temporal width of \(\sigma_t\), the spectral ACF exhibits a Gaussian peak with a width given by \(\sigma_{\text{self}} = \frac{1}{2\pi\sigma_t}\). In the specific case where the spectral ACF width \(\sigma_{\text{self}}\) is smaller than the scintillation bandwidths, the spectral ACF of a Gaussian pulse interacting with two unresolved screens shows an additional Gaussian curve above a correlation of three, as illustrated in \cref{fig:Case1_Gaussian_ACF}. For a one-to-one comparison, the spectral ACF when the intrinsic pulse is delta function is shown in the inset of \cref{fig:Case1.1 - ACF}, here the self-noise is absent. If the self-noise width is comparable to the \(\nu_s\) of one of the screens, it results in a composite profile in the spectral ACF, and the product terms including \(\text{ACF}_{\text{MW}} \times \text{ACF}_{\text{host}} \times \text{ACF}_{\text{self}}\) in \cref{Eq:TwoScreenACF} inhibits the measurement of scintillation parameters. If one reduces the intrinsic pulse width, the self-noise width increases, and in the limit of the pulse approaching a delta function, the self-noise disappears.

For an observer to distinguish between self-noise and scattering, one can divide the pulse into subbands and observe the evolution of the individual profiles in the spectral ACF. Scattering produces Lorentzian profiles whose half-width at half-maximum (HWHM) scales with frequency as \(\nu_{\text{s}} \propto \nu_{\text{obs}}^4\), whereas the self-noise width is frequency independent, assuming that the burst structure is also frequency independent.

To illustrate the effect of self-noise in the ACF, we use a temporally wide Gaussian intrinsic pulse, such that the self-noise feature in the ACF has a width much smaller than the scintillation bandwidths of the scattering screens and remains clearly distinguishable. The dynamic spectrum of this case generated by resolving broad scintillation features in frequency while preserving the temporal profile of the pulse is shown in the top-left plot of \cref{fig:Dynspec_combined}. In the dynamic spectrum we observe a constant scintillation pattern across the pulse time profile, characteristic of a pulse encountering an unresolved two-screen system.  

For quantitative analysis, we compute the full-spectrum ACF (\cref{fig:Case1_Gaussian_ACF}). Due to the distinct scales of scintillation and self-noise, individual profiles in the ACF can be separated through thresholding. The scintillation bandwidth and modulation index obtained from Lorentzian fits remain the same as quoted in \cref{sec:1.1} and is in good agreement with the injected values. However, the modulation index estimated using \cref{eq:mod_1} exceeds \(\sqrt{3}\), due to additional self-noise modulation. Therefore, we argue that when analyzing the full-spectrum ACF, the peak correlation of narrow scintillation, as determined from Lorentzian fits, should be used as a more reliable measurement of the total modulation from the two-screen system.

\subsection{Screens that just resolve each other}
\label{subsec:Justresolving}

FRB 121102A is utilized as the source to illustrate this case, with modifications made to the screen distances and screen sizes. To demonstrate the qualitative features, we examine the simulated dynamic spectra produced using a Gaussian intrinsic pulse (\cref{fig:Dynspec_combined}, Bottom-left plot). The channelization is chosen such that it resolves frequency scales equal to and larger than expected from the MW scintillation. The scintillation pattern remains almost constant across the pulse, showing no drastic difference in the pulse dynamic spectrum compared to unresolved screens. We also note some smearing in the scintillation pattern, suggesting that resolving effects start being visible at RP$=1$.

For the quantitative analysis, we examine the spectral ACF in \cref{fig:Case2 - ACF}. The fit to MW scintillation gives a HWHM of \(\nu_{\text{s,MW}} \approx 200\) kHz, compared to the injected value of 160 kHz (Table \ref{tab:RP=1 screen param} in the Appendix). The modulation index of the MW screen is \(m_{\text{MW}} \approx 0.83\), which is less than 1. The slight broadening and decrease in modulation is an effect of the screens partially resolving each other. Meanwhile, the host screen scintillation has a fit value of \(\nu_{\text{s,MW}} \approx 1\) kHz, which is in complete agreement with the injected value. The total modulation index for the two-screen system is 1.54.
This decrease in modulation index from \(\sqrt{3}\) agrees with our theoretical expectation for a partially resolving two-screen system. Furthermore, the ACF also shows that only the broad-scale scintillation is affected by the screens resolving one another. 

\subsection{Resolving screens}

\label{subsec:resolving_Screens}

To illustrate this case, we consider a closer FRB, FRB 20180916B, localized to a spiral galaxy at a redshift of \(z_{\text{FRB}} \approx 0.0337\) \citep{2020Natur.577..190M}, with a corresponding angular diameter distance of \(D_{\text{FRB}} \approx 138~\text{Mpc}\). The simulated pulse dynamic spectrum, shown in \cref{fig:Dynspec_combined} (bottom-right plot), is channelized to resolve the broad-scale MW screen scintillation while maintaining sufficient time resolution to observe the scattering tail. The presence of small secondary peaks in the time profile, instead of a smooth scattering tail, can be attributed to the use of a finite number of images in the screen. An intriguing observation from the dynamic spectrum is the variability of scintillation throughout the pulse.

\cref{fig:Case3 - ACF} shows the full-spectrum ACF of the pulse. The fit to host screen scintillation gives \(\nu_{\text{s,host}} \approx 0.2\) kHz, which agrees with the injected value. On the other hand, the Lorentzian fit to the broad scintillation gives a HWHM of \(\nu_{\text{s,MW}} \approx 0.8\) MHz, which is much greater than the injected value of 0.16 MHz. Furthermore, the total two-screen modulation from the average spectrum is \(m_{\text{tot}} \approx 1.3\), determined from the zero-lag of the ACF, while the modulation index of broad scintillation determined from the Lorentzian fit to the broad scales (green), is as low as (\(m_{\text{MW}}\approx 0.2\)) due to quenching. The broadening of only the MW scintillation bandwidth, as well as the reduction in the MW screen modulation index, solidify our observation that only broad scale scintillation is affected by resolving.

The dynamic spectrum for this case (bottom-right plot in \cref{fig:Dynspec_combined}) exhibits a scintillation pattern that varies across the pulse time profile. The dotted lines in the plot compare the injected MW scintillation bandwidth with the bandwidth measured by fitting the full spectral ACF. This comparison highlights that the fitted scale does not fully capture the range of scintillation sizes present. This discrepancy arises because the scattering screens are no longer point-like relative to each other, resulting in scintillation patterns at different scales. The full-spectrum ACF introduces an averaging effect, emphasizing patterns of similar magnitude and yielding an average scintillation bandwidth, further discussion on this effect can be found in \cref{sec: Quenching of scintillation}. To explore the evolution of scintillation over the pulse duration, we take slices of the pulse dynamic spectrum in time and measure the modulation index and scintillation bandwidth of the spectrum at each time bin. The intra-pulse evolution of $\nu_{\text{s,MW}}$ is shown in \cref{fig:Case3_scint evol}, where we observe a linear increase. The $m_{\text{MW}}$ evolution across the pulse profile for this case is illustrated with red triangles in \cref{fig: MOD-EVOL}. The intra-pulse decreasing trend of $m_{\text{MW}}$ observed in our simulation is consistent with the observational results for FRB 20201124A reported by \citet{sammons2023two}, who noted a similar decline in the modulation index across the pulse. A detailed comparison of observables for bursts at different resolution regimes of two-screen scattering is provided in \cref{sec: Quenching of scintillation}.

\begin{figure}
    \centering
    \includegraphics[width=\linewidth]{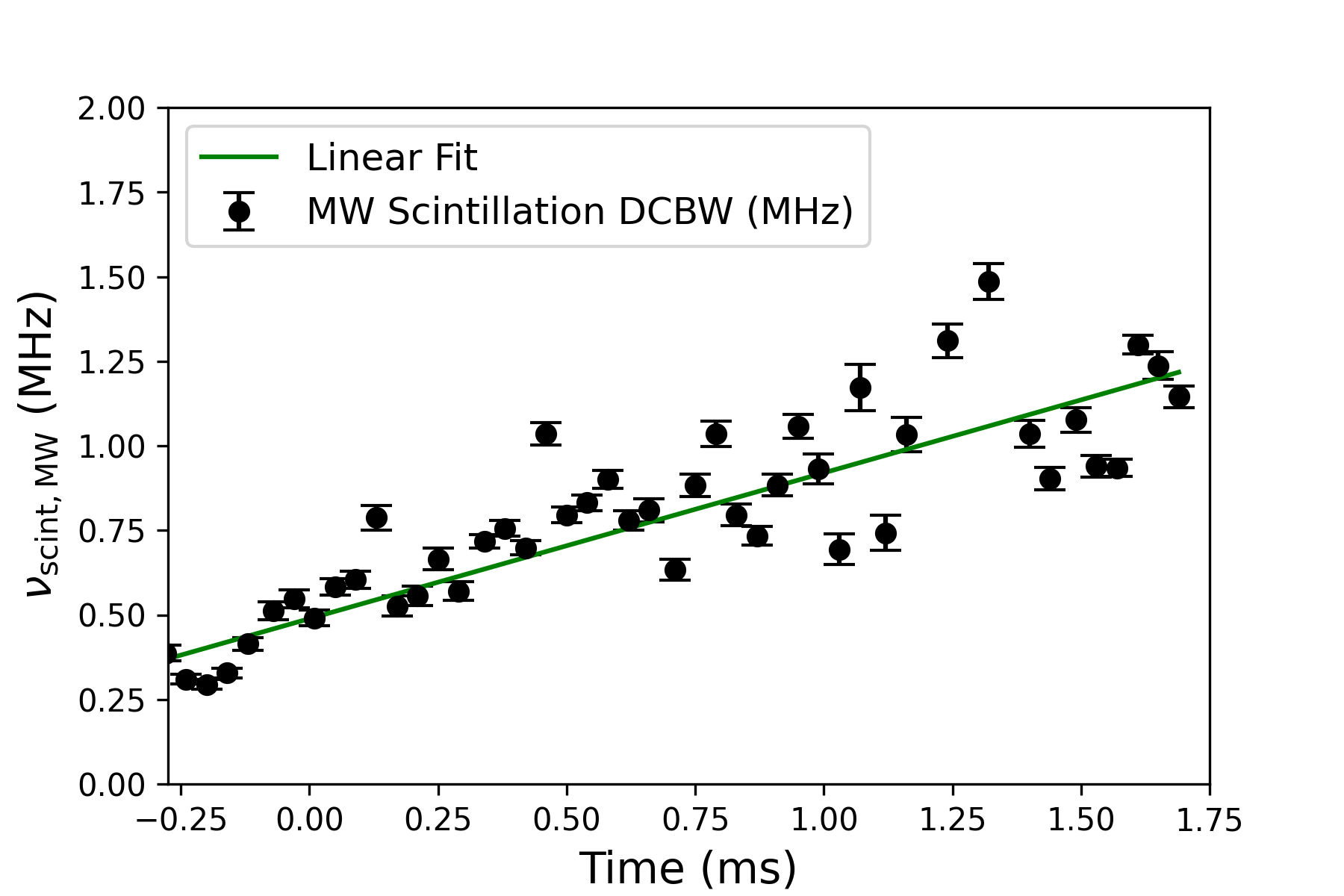}
    \caption{MW scintillation bandwidth as a function of time bins along the burst, derived from the dynamic spectrum of the completely resolving screens simulation shown in the bottom-right plot of \cref{fig:Dynspec_combined}. The time bin interval corresponds to the time resolution of the pulse dynamic spectrum. Each point (black dot) corresponds to the scintillation bandwidth measured from Lorentzian fits to scintillation at each time bin, with error bars indicating the fit error. The linear fit to the black points illustrates the increase in scintillation bandwidth over the pulse duration. The systematic error, due to the finite sampling of the screen, results in the scattering of data points around the linear fit, which is not captured by the fitting error.}
    \label{fig:Case3_scint evol}
\end{figure}

\begin{figure}
    \centering
    \includegraphics[width=1\linewidth]{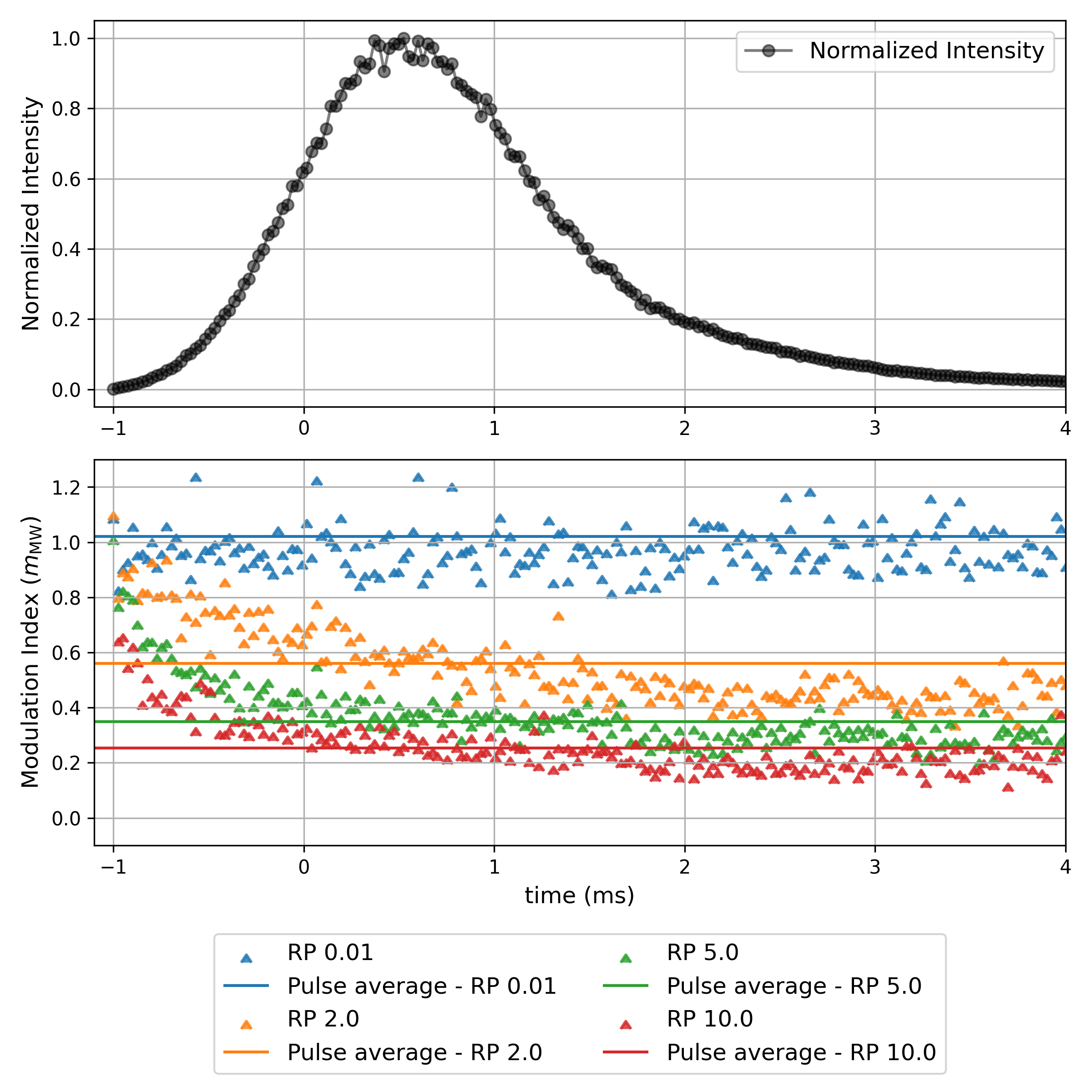}
    \caption{Intra-pulse evolution of modulation index. \textbf{Top:} Channelized pulse intensity averaged over frequency. \textbf{Bottom:} Colored triangles represent the modulation index of the spectrum at each time bin, with each color corresponding to results from a system with different RP values. The solid lines represent the modulation index measurements from the pulse averaged spectrum of these systems. The channelization provides sufficient frequency resolution to resolve all the MW's scintillation. $\text{RP}<1$ when the screens do not resolve each other and $\text{RP}>1$ when screens resolve each other.} 
    \label{fig: MOD-EVOL}
\end{figure}

\section{Implications and Applications}

In this section, we compare systems with different RP, discuss resolutions effects in one-dimensional screens and the frequency dependent evolution of RP and its impact on scintillation bandwidth. Additionally, we also review existing formulas used to place upper limits on the distance between the FRB source and the host scattering screen. 
\label{Sec: Discussion}

\subsection{Quenching of scintillation and other comparisons}
\label{sec: Quenching of scintillation}

In Section~\ref{subsec:resolving_Screens} we showed that when the MW screen resolves the host screen, the observed the MW scintillation bandwidth is larger than and modulation index lower than what is expected from the unresolved or single-screen case. In order to further illustrate the observational signatures of the transition from unresolved to resolved,  we define several two-screen systems with different resolution powers in which all screen parameters are held fixed except for the distance between the MW screen and observer and the size of the MW screen. These are varied in such a way that the injected MW delay (or the MW scintillation bandwidth) remains the same facilitating one to one comparisons. 

\begin{figure}
    \centering
    \includegraphics[width=1\linewidth]{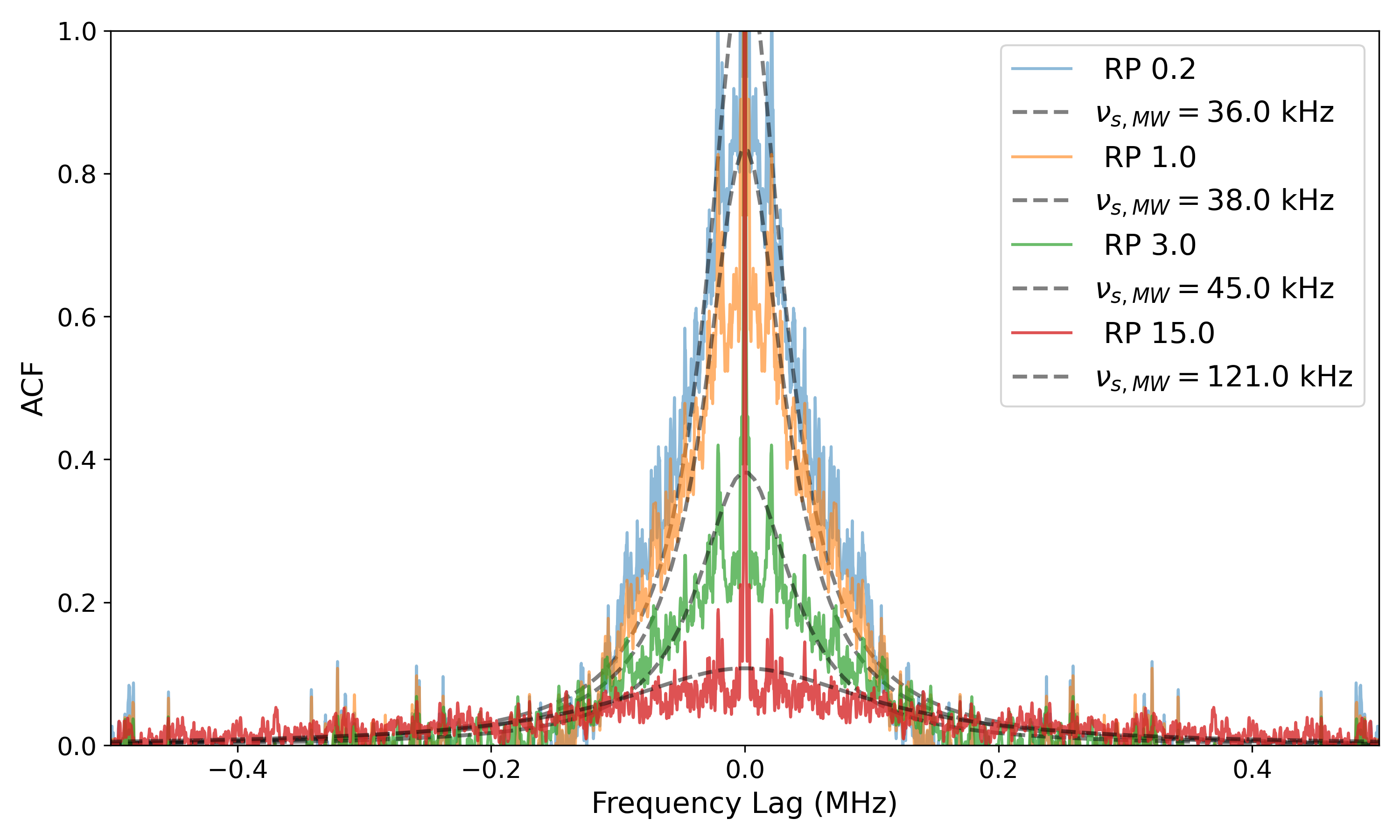}
    \caption{ACF evolution of broad scintillation with the resolution power (RP) of the system. The plot represents a magnified view of the data along the x-axis, spanning the frequency lag range from -0.4 MHz to 0.4 MHz. Colored lines represent the spectral ACF of two-screen systems with different RP values, while the dashed curves are Lorentzian fits to the ACFs. Systems with RP values in the range 0.2 to 15 are plotted to illustrate the gradual quenching of broad scintillation. The injected MW scintillation bandwidth is 30 kHz for all systems, facilitating a one to one comparison of the ACFs.}
    \label{fig: ACF-RP}
\end{figure}

To compare the full-spectrum ACFs of a pulse encountering these two-screen systems at different resolving regimes, we simulate systems with RP values of \(0.2\), \(1.0\), \(3.0\), and \(15.0\) and fit only the broad scintillation in each ACF with a Lorentzian. 
The results are shown in \cref{fig: ACF-RP}. Comparing the measured scintillation bandwidths with expected value of 36 kHz for the broad MW scintillation in the single screen case, it becomes clear that as the screens start to resolve each other, the broad scintillation becomes both broadened and damped simultaneously. For example, the unresolved system (blue curve, RP=0.2) shows a peak MW scintillation correlation of 1 while the just resolved system (orange curve, RP=1) shows a peak correlation of 0.8. As the two screens resolve each other more (green, RP=3), the peak correlation reduces to 0.4, and a further increase in RP causes further dampening of the correlation. 

As previously demonstrated in the fully resolved screen case (\cref{subsec:resolving_Screens}), 
a full-spectrum ACF does not capture any intra-pulse evolution of the scintillation.
A one-to-one comparison of the intra-pulse evolution of the MW modulation index in simulations of two-screen systems with RP values of \(0.01\), \(2.0\), \(5.0\), and \(10.0\) is shown in \cref{fig: MOD-EVOL}. In the plot the modulation index at each time bin of the dynamic spectra (triangles) is measured from the Lorentzian fits to the respective spectrum. We observe that for unresolved screens, $m_{\text{MW}}$ remains close to 1 throughout the pulse. In contrast, for screens that resolve each other, $m_{\text{MW}}$ starts with a value below 1 at the beginning of the pulse and decreases over time. For comparison, the solid lines show the modulation index computed from the pulse-averaged spectrum, where the dynamic spectrum is averaged over the duration of the pulse. Both \cref{subsec:resolving_Screens} and \cref{fig: ACF-RP} highlight that the effects of resolution on observables are strongly dependent on the RP of the screen system, and broad scale scintillation persists unless the screens resolve each other completely.

When we measure the scintillation bandwidth and modulation index from the autocorrelation function (ACF) of the full-spectrum, we capture the combined effect of all the scintillation scales present during the pulse. In the case of resolved screens, this combined effect is equivalent to summing multiple scintillation components with a range of bandwidths, much like adding Lorentzian profiles of different widths. The widths range from the injected value to larger values due to broadening over the pulse time. The superposition of these Lorentzians results in a broader composite profile compared to the injected value, explaining the observed broadening of the scintillation bandwidth compared to the injected value. Furthermore, since the scintillation patterns at each time bin are less correlated with each other, the total modulation index is reduced due to the averaging effect of uncorrelated fluctuations.

One way to visualize the change in scintillation patterns over the pulse duration is by imagining the screens as interferometers and considering which light paths contribute to each time bin in the dynamic spectrum. 
In our scenario, the scattering tail is governed by the host screen, which produces larger delays, and the scintillation is governed by the MW screen, which produces smaller delays. 
In the first time bin, light travels along the shortest path, passing through the central region of the host screen. The diameter of this central region is small enough to be unresolved by the entire MW screen, resulting in the scintillation with the smallest bandwidth over the pulse duration. Successive time bins originate from rings with increasing radii in the host screen, which become progressively more resolved by the MW screen, leading to an increase in RP and scintillation bandwidth over the pulse duration.

Once the MW screen resolves the host screen, the effective baseline within the MW screen over which the phases interfere constructively becomes smaller than the MW screen baseline in the unresolved regime. Light from a ring in the host screen, with a diameter large enough to be resolved by the longest baseline in the MW screen, produces numerous small coherent patches in the MW screen. As the ring continues expanding across the pulse time, the coherent patch size in the MW screen decreases. The patch size dictates the scintillation bandwidth, causing the intra-pulse increase in scintillation bandwidth. Additionally, each of these patches are spatially separated. The phase fluctuations in the MW screen are random and uncorrelated over these separations. As a result, the signals from these patches interfere at the observer with varying phases. This results in a decrease in the overall coherence and, consequently, a reduction in the correlation of the scintillation pattern. This picture provides a qualitative explanation for the increase in scintillation bandwidth and the decrease in modulation index across the time axis of the pulse. Based on this picture, the observed linearly increasing intra-pulse evolution of scintillation bandwidth in \cref{fig:Case3_scint evol} corresponds to a quadratic decrease in the size of coherent patches in the MW screen over the pulse duration.

\subsection{Locating the host galaxy screen }
\label{sec: host screen distance}

\begin{table*}[]
    \centering
    \renewcommand{\arraystretch}{2.0}
    \caption{Comparison of formulas proposed to constrain the distance between FRB source and host galaxy screen.}
    \label{tab:formula comparision}
    \begin{tabular}{c|lccc}
    \hline\hline
    Study & Distance estimation formula & $D_{\text{h,FRB 20190520B}}$ & $D_{\text{h,FRB 20201124A}}$ & $D_{\text{h,FRB 20221022A}}$ \\
    \hline
    \citet{main2022scintillation} & $D_{\text{h,FRB}}D_{\text{MW}} \lesssim \frac{\pi D_{\text{FRB}}^2}{2 \nu^2}\frac{\nu_{\text{s,MW}}}{\tau_{\text{s,h}}}$ & $\lesssim 1.1$\,kpc & $\lesssim 34$\,kpc & $\lesssim 135$\,kpc \\
    \makecell{\citet{2022ApJ...931...87O} \\ \citet{2025Natur.637...48N}} & $D_{\text{h,FRB}}D_{\text{MW}} \lesssim \frac{D_{\text{FRB}}^2}{2 \pi \nu^2} \frac{\nu_{\text{s,MW}}}{ \tau_{\text{s,h}} }$ & $\lesssim 0.11$\,kpc & $\lesssim 3.4$\,kpc & $\lesssim 14$\,kpc \\
    \citet{sammons2023two} & $D_{\text{h,FRB}}D_{\text{MW}} \approx \frac{D_{\text{FRB}}^2}{2 \pi \nu^2 (1+z_{\text{FRB}})}  \frac{\nu_{\text{s,MW}}}{\left(m_{\text{MW}}\right)^2 \tau_{\text{s,h}}}$ & ($\approx 0.088$\,kpc) & $\approx 9.0$\,kpc & $\approx 22$\,kpc \\
    This work & $D_{\text{h,FRB}}D_{\text{MW}} \lesssim \frac{(1 + z_{\text{FRB}}) D_{\text{FRB}}^2}{8 \pi \nu^2 } \frac{\nu_{\text{s,MW}}}{m_{\text{MW}} \tau_{\text{s,h}}}$ & ($\lesssim 0.034$\,kpc) & $\lesssim 1.6$\,kpc & $\lesssim 4.5$\,kpc \\
    \hline
    \end{tabular}
    
\end{table*}

The observation of scintillation and scattering from different screens means that the two screens need to be placed such that the scintillation is not quenched completely. Thus, information about the distance between the farther screen and the FRB is obtained. \citet{masui2015dense}, \citet{main2022scintillation} and \citet{2022ApJ...931...87O} have used this fact to place upper limits for this distance to observed FRBs. They invoke the condition that the angular resolution $\theta_\text{res}$ provided by the the size of the MW screen needs to be larger than the angular size $\Theta_\text{size}$ of the host screen so that scintillation is not quenched. \citet{sammons2023two} expanded this argument by including the modulation index in order to arrive at an estimate rather than an upper limit. Here, we follow the same idea and point out some needed modifications. All four proposed distance estimation formulas are summarized in \cref{tab:formula comparision}.

We adopt an assumption made by all previous authors that the distance of the FRB is much larger than the distance of the screens to their corresponding host galaxies, such that $D_{\text{MW},\text{host}}=D_\text{host}=D_\text{src}$, and all of these variables can be replaced by the distance $D_{\text{FRB}}$ of the FRB. In addition, a condition of such an estimation is the finding that scattering time $\tau_s$ and scintillation bandwidth $\nu_s$ belong to clearly distuingishable scattering screens. In this case, the host screen's scattering time can be taken from \cref{Eq:tau_s_host} and is independent of the resolution power:
\begin{equation}
    \tau_{s,\text{h}} = \frac{1+z_\text{FRB}}{2c} \, \frac{D_\text{FRB}^2}{D_\text{h,FRB}} \,\theta_\text{L,h}^{2} \, . \label{Eq:tau_s_for_formula}
\end{equation}
Here, we deviate from \citet{sammons2023two} due to the correction to \citet{macquart2013temporal}, as discussed in \cref{Sec:CosmoScatteringTheory}. \citet{main2022scintillation} and \citet{2022ApJ...931...87O} used low-redshift approximations of \cref{Eq:tau_s_for_formula}.

Many prior works have been based on the assumption that the scintillation bandwidth is independent of the resolution power, i.e.~ that \cref{Eq:nu_s_MW} is valid:
\begin{equation}
    \nu_{s,\text{MW}} = \frac{c}{\pi D_\text{MW}} \,\theta_\text{L,MW}^{-2} \, .
\end{equation}
The broadening effect that we observe in our simulations makes this assumption invalid. We follow \citet{2025Natur.637...48N} and use Equation 46 of \citet{1998ApJ...505..928G} to describe this effect. Although derived for an incoherent extended source, we found it to be in agreement with our simulations. In our notation, it reads
\begin{equation}
    \nu_{s,\text{MW}} = \frac{c}{\pi D_\text{MW}} \,\theta_\text{L,MW}^{-2} \sqrt{1+\frac{\pi^2}{4^3}\text{RP}^2} \, .\label{Eq:nu_s_for_formula}
\end{equation}

The dependence of the modulation index on the resolution of a source was also already known by \citet{1998ApJ...505..928G}, and we again follow \citet{2025Natur.637...48N} in using it also for resolving a screen, which is supported by our simulations as shown in \cref{fig:fit_vs_theory_over_RP}:
\begin{equation}
    m_{\text{MW}} = 1 / \sqrt{1+\frac{\pi^2}{4^3}\text{RP}^2} \, . \label{Eq:m_for_formula}
\end{equation}
Conveniently, this result is exactly the inverse of the broadening factor in \cref{Eq:nu_s_for_formula}. Hence, we can restore the ansatz of \citet{sammons2023two} by canceling the broadening in $\nu_{s,\text{MW}}$ with an extra factor of $m_{\text{MW}}$. They take a different formula for the modulation index that was given by \citet{narayan1992physics}:
\begin{equation}
     m_{\text{MW}} = \frac{\lambda}{2\pi\theta_{\text{L,MW}}\theta_{\text{L,h}}D_{\text{MW}}} = \frac{8}{\pi \text{RP}} \, .
     \label{mod_index_1}
\end{equation}
We note that this is the limit of \cref{Eq:m_for_formula} for high resolution powers. Thus, $m_{\text{MW}}$ is lesser than or equal to this value.

Combining \cref{Eq:nu_s_for_formula,Eq:tau_s_for_formula,Eq:m_for_formula} and using \cref{mod_index_1} as an upper limit, we obtain
\begin{equation}
    D_{\text{h,FRB}}D_{\text{MW}} \lesssim \frac{(1 + z_{\text{FRB}}) D_{\text{FRB}}^2}{8 \pi \nu^2 } \frac{\nu_{\text{s,MW}}}{m_{\text{MW}} \tau_{\text{s,h}}} \, . \label{Sachin 2024 formula}
\end{equation}

For a localized FRB the angular diameter distance $D_{\text{FRB}}$ to the host galaxy can be calculated from its redshift $z_\text{FRB}$ using a model of cosmology. Prior information on the distance to the MW scattering region can be found using a Galactic electron density model like NE2001 by \cite{cordes2002ne2001} or YMW16 by \citet{yao2017new}. However, this approach returns large uncertainties and requires that the FRB sightline passes through a studied region. Better results are expected if the screen's location can be solved by monitoring the scintillation of an active repeater over the course of a year \citep{2023MNRAS.522L..36M,wu2024scintillation}. Another possibility to determine the distance to the MW screen would be to measure the angular size of the scattering disk $\theta_\text{L,MW}$ directly. This method has been demonstrated for FRB 121102A in \cite{Ocker2021ApJ...911..102O} using VLBI measurements of angular broadening from \cite{2017ApJ...834L...8M}.

In \cref{tab:formula comparision}, we listed the constraints resulting from the proposed formulas for the FRBs 20190520B, 20201124A, and 20221022A using the same measurements and assumptions as were reported in the respective studies by \citet{2022ApJ...931...87O}, \citet{sammons2023two}, and \citet{2025Natur.637...48N}\footnote{See \cref{sec: Simulation parameters} for the values used.}. In the case of FRB 20190520B the modulation index is not known, and a value of 1 was used instead. Thus, the results for formulas using the modulation index may be too low. 

While following an equivalent logic, the formula by \citet{main2022scintillation} differs by a factor of $\pi^2$ from the formula derived by \citet{2022ApJ...931...87O}, which was also used by \citet{2025Natur.637...48N}. This significant difference is a result of different definitions of the threshold between resolved and unresolved, which is a smooth transition in reality. For FRBs in the local universe and in the absence of resolution effects ($m_{\text{MW}}\approx1$), our proposed formula also deviates only by a constant factor leading to even tighter constraints, which do not rely on such a threshold. Our findings show that the formula derived by \citet{sammons2023two} should also be interpreted as an upper limit because they assumed different evolutions of the scintillation bandwidth and modulation index with resolution.

\begin{figure}
    \centering    
    \includegraphics[width=\linewidth]{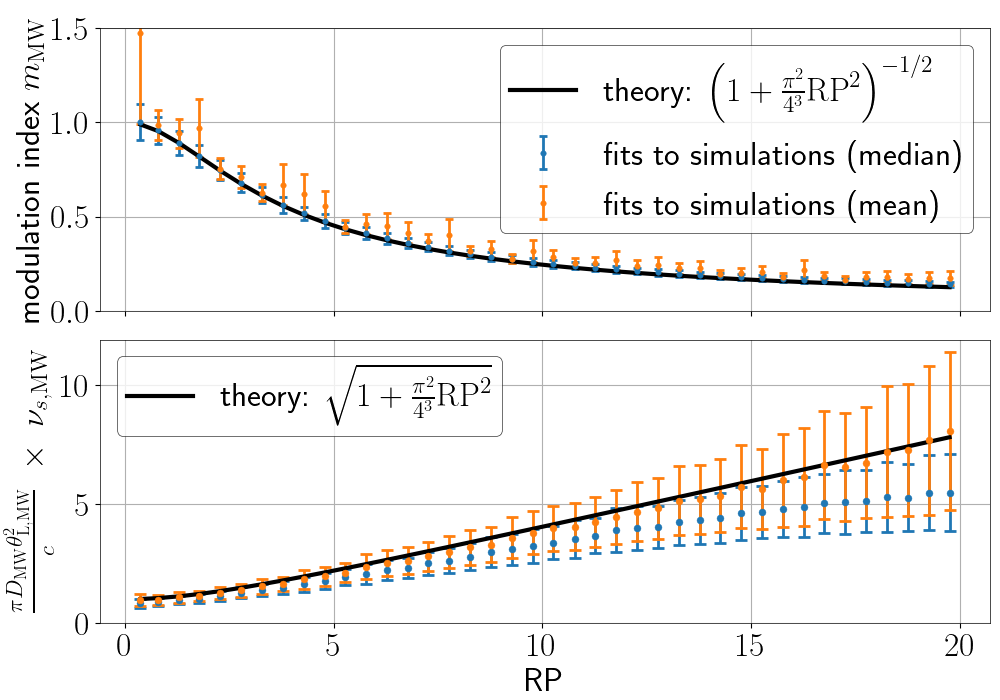}
    \caption{Effect of increasing evolution on modulation index and scintillation bandwidth. The black line shows the theoretical relations used here. The data points show the median and mean of all successful fits to simulation within the corresponding RP bins. The errorbars correspond to the root median square deviation from that value.}
    \label{fig:fit_vs_theory_over_RP}
\end{figure}

\begin{figure}
    \centering    
    \includegraphics[width=\linewidth]{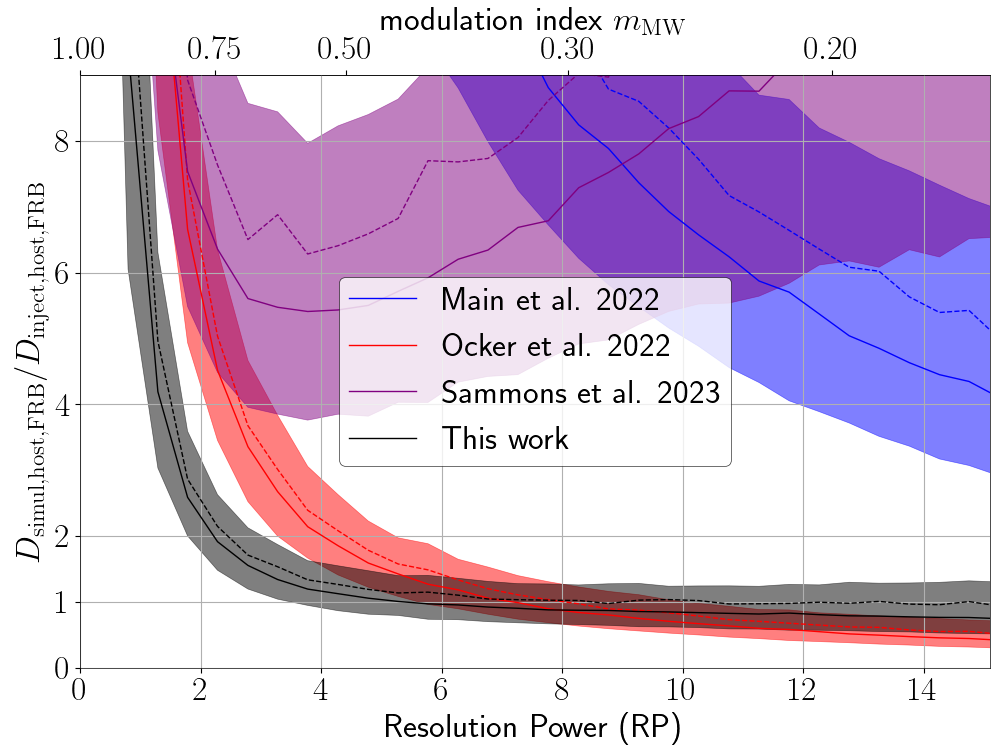}
    \caption{Comparison of formulas put forward to place upper limits on the host galaxy screen distance from the FRB. For bins of resolution power, the median is shown as a solid line, the mean is shown as a dashed line and the shaded regions correspond to the central 68\% of the data distributions. The corresponding modulation index from \cref{Eq:m_for_formula} is shown at the top. The upper limits by \citet{main2022scintillation} and \citet{2022ApJ...931...87O} were derived for $m_\text{MW}\approx 1$ and are shown here outside of that regime. }
    \label{fig:Dist lim comp}
\end{figure}

To compare the effectiveness of the proposed formulas in estimating the distance between the FRB source and its host scattering screen within the boundaries of our modeling, we ran simulations on randomized systems, whose parameter ranges are listed in \cref{sec: Simulation parameters}. The modulation index and the scintillation bandwidth are fitted as described in \cref{Sec: FRB_Scintillator} and a comparison between theory and simulations is shown in \cref{fig:fit_vs_theory_over_RP}. For the parameters $D_{\text{MW}}$ and $\tau_{\text{s,h}}$ the injected values are used because any estimation would require a realistic modeling of the burst structure and its dispersion. The resulting ratios of the upper limit of $D_{\text{h,FRB}}$ to its injected value are binned according to the resolution powers. Within these bins, the median and the root median square deviation from the median are computed. 

The results are shown in \cref{fig:Dist lim comp}. The use of the median is necessary because of significant outliers. The simulations reveal significant random as well as systematic deviations. Since the simulations contain a realistic number of individual images, these deviations are to be expected in real data rather than being a shortcoming of numerical simulations. Individual realisations of the Gaussian random distributions of variables have a large impact on the result. Furthermore, simple fits of Lorentzian functions seem to be systematically biased, such that improved estimation methods should be considered in the future. At RP values below 4, all formulas become inequalities at best, with \cref{Sachin 2024 formula} providing the constraints closest to the real value. As can be seen in \cref{fig:fit_vs_theory_over_RP}, the results of fits show significant scatter around the theoretical value and follow statistical distributions where the mean and median are different, such that individual measurements should be treated with sufficient caution.

\subsection{Two 1D screens}
\label{sec:1-D screens}

Now we consider the case where the shape of both the host galaxy and MW scattering screens are highly elongated in the plane of the sky (i.e. "one-dimensional") (1D) by simulating a straight line of images instead of the two-dimensional (2D) screen. Geometrically this could be either an intrinsically 1D filament or 2D thin sheet viewed edge on. A highly elogated screen has been unambiguously observed for PSR B0834+06 by \citet{Brisken_2010}, and  \citet{2025ApJ...980...80S} compiled a list of the growing number of scintillating pulsars with measured 1D screen orientations. Notably, one-dimensional scattering in the sky plane has only been confirmed by making use of advanced scintillometry techniques; this geometry was unknown when only the three observables of scintillation bandwidth, scattering time, and modulation index were accessible. Since scintillation studies of FRBs are mostly limited to these observables, the geometry of the scattering screen remains ambiguous in most cases. The growing list of 1D screens found for pulsars in the MW implies that not all FRBs undergo scattering from 2D screens.

Now, the angular coordinates of the images have to be treated as 2-D vectors in \cref{Sec:CosmoScatteringTheory} such that the mixed term in \cref{Eq:tau_eff} depends on the relative orientation of two screens. For a perpendicular orientation, this term vanishes, and the screens do not resolve each other regardless of their location and size. The model defined in \cref{Sec:Models} has to be adapted such that the random variables are only Gaussian along the line of the screen and zero otherwise. In this section we briefly look into two cases:
\begin{enumerate}
    \item Two 1D screens that are aligned parallel to each other
    \item Two 1D screens that are aligned perpendicular to each other.
\end{enumerate}

Similar to before we consider the screen in the host galaxy ISM to contribute the highest delays and the screen in the MW ISM to produce the broad scintillation over frequency. For both cases the initial parameters of the simulations remain the same with the only difference being the screen orientation. The resolution power of the screen system is 12.4, for which a 2D two-screen system shows significant dampening and broadening of scintillation over frequency as discussed in \cref{sec: Quenching of scintillation}. The simulations are repeated 20 times with different image seeds to account for systematic errors. The mean value of the observables from each simulation is presented as the measured observable in this analysis. 

\begin{figure*}
    \centering
    \includegraphics[width=1\linewidth]{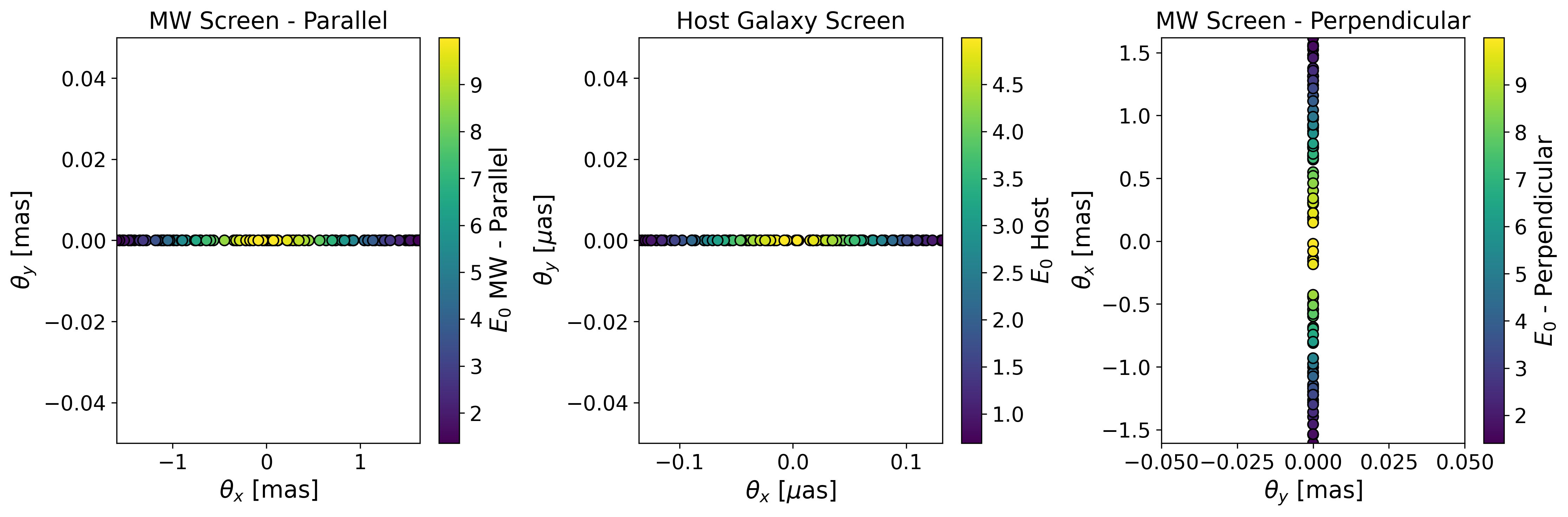}
    \includegraphics[width=0.8\linewidth]{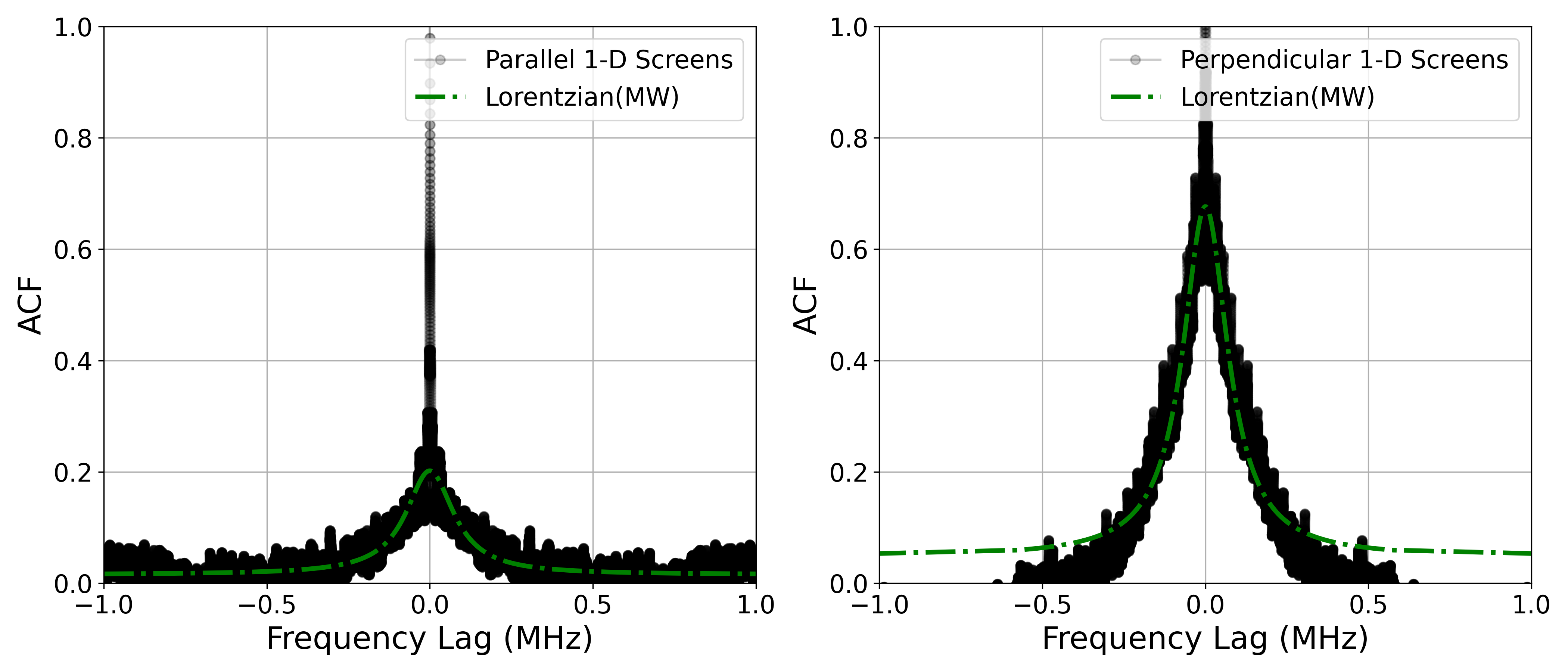}
    
    \caption{\textbf{Top:} The left and mid panels form a parallel-aligned screen system, while the right and mid panels form a perpendicular screen system. Both systems share the same screen parameters, with a RP=12.4, calculated using \cref{eq:RP_def1}. \textbf{Bottom:} Spectral ACF shwoing the central broad scintillation from the MW screen. \textbf{Left:} The ACF of the aligned system, where the correlation of broad scintillation is significantly less than 1, indicating quenching of scintillation. \textbf{Right:}  The ACF of the perpendicular system, with the broad scintillation closer to 1 and no visible signs of quenching.} 
    \label{fig:two_1-D_comparision}
\end{figure*}

The image distributions and corresponding ACFs from a single realization of two 1-D screens with perpendicular and parallel configurations, selected from 20 simulations, are shown in \cref{fig:two_1-D_comparision}. The ACF of parallel screen system shows quenching while the perpendicular system shows no sign of quenching, demonstrating that screen alignment plays a major role in quenching of scintillation in 1D screen systems. This is because RP is the ratio of the angular size of the second screen at the first screen and the angular resolution of the first screen. Using an interferometer as an analogy, a straight line of image points resembles a straight baseline of antennas, which has a high resolution only in one direction. Similarly, a 1D screen can only resolve structures parallel to the screen's orientation.  

For two 1D screens that are not aligned, the formulation of RP should be corrected by the projection of one screen onto the other. As a consequence, the distance limit formula defined in \cref{sec: host screen distance} to find the distance between the host galaxy screen and FRB is no longer valid. In \cref{tab:distance formula_2D-1D} we compare the injected values and calculated values of \(D_{\text{host,src}}\), using scintillation bandwidth and modulation index measurements across 20 simulations, for different screen configurations: two 2D screens, two parallel 1D screens, and two perpendicular 1D screens. \cref{tab:distance formula_2D-1D} shows that the distance limit formula predicts $D_{\text{host,src}}$ far from its input value for perpendicular 1D screens. For partially aligned systems, $D_{\text{host,src}}$ is predicted close to the input value but this may not be the case for all possible parameters.

\begin{table}
    \centering
    \renewcommand{\arraystretch}{1.2}
    \begin{tabular}{l|ccc}
        \hline
        Two-screen geometry & $\nu_{\text{s,MW}}$  & $m_{\text{MW}}$ & $D_{\text{host,src}}$   \\
        \hline
       2D screens & 190\,kHz  & 0.28 & 4.6\,kpc \\
        \hline
       Parallel 1D screens & 320\,kHz & 0.42 & 5.1\,kpc  \\
        \hline
        Perpendicular 1D screens& 49\,kHz & 0.7 & 0.5\,kpc \\
        \hline
    \end{tabular}
    \caption{A comparison of host screen distance from the FRB source ($D_{\text{host,src}}$) deduced using \cref{Sachin 2024 formula}, between two 1D screens and 2D screens. The table presents the mean values of the measured observables - modulation index ($m_{\text{MW}}$) and scintillation bandwidth of the broad scintillation ($\nu_{\text{s,MW}}$)- from 20 simulations. Fit errors and systematic errors are less than 10 percent and hence not presented here. Using these observabels we measure $D_{\text{host,src}}$ via \cref{Sachin 2024 formula}. Comparison with the injected value of $D_{\text{host,src}} = 5$ kpc indicates that the resolution arguments and distance-limit formulas are only valid when the screens are 2D or well aligned.}
    \label{tab:distance formula_2D-1D}
\end{table}

\subsection{Observable trends in broadband studies}

In \cref{Eq:nu-2scaling} we have assumed that the width of the distribution of scattering angles follows $\nu^{-2}$. In the unresolved case this means that the scattering delay follows $\nu^{-4}$ and the scintillation bandwidth follows $\nu^{4}$. Significant deviations from this law can be indicative of screens resolving each other with the clearest signal being a flattening towards lower frequencies, where the resolved case is more likely due to larger angles. A similar scenario arises when a screen resolves the emission region of a fast radio burst (FRB) or a pulsar. Broadband observations can capture the transition from a two-screen system where the screens do not resolve each other to one where they do, or from a screen perceiving the emission region as point-like to recognizing it as extended and resolved. By analyzing the behavior of the scintillation bandwidth across different observing frequencies, one can detect and distinguish between these scenarios.

\subsubsection{Screen resolving another screen}

\begin{figure}
    \centering    
    \includegraphics[width=\linewidth]{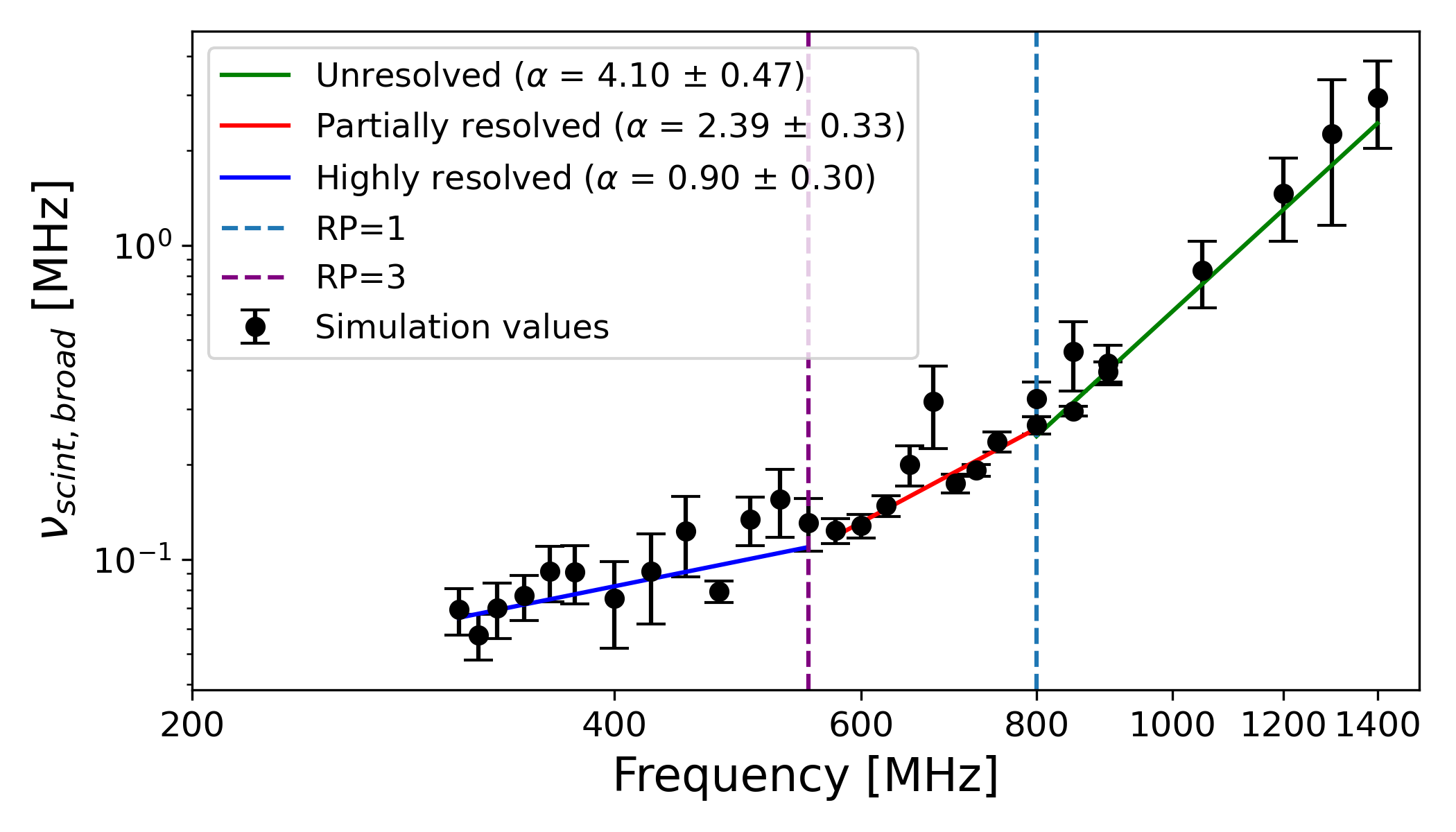}
    \caption{Scintillation bandwidth evolution with frequency for a two-screen system. Black dots represent mean values from 30 simulations with error bars indicating the standard error. The RP=1 and RP=3 lines separate the regions where screens are in the unsolved, partially resolved, and highly resolved regimes. Fits to the simulation values show a broken power-law distribution (\(\nu_{\text{s,MW}} \propto \nu^{\alpha}\)): \(\alpha \approx 2\) in the resolved region and \(\alpha \approx 4\) in the unresolved region. }
    \label{fig:scint BW - freq}
\end{figure}

We simulated scintillation bandwidths according to images placed on two 2D scattering screens whose size follows $\theta_L\propto \nu^{-2}$. To account for statistical variations from simulation to simulation, 30 simulations are produced at a given central frequency with different image seedings. This was repeated at several frequencies. In addition, to avoid self-noise effects in the spectral ACF, a delta function was used as the intrinsic pulse.

The results are shown in \cref{fig:scint BW - freq}. We use power laws to fit different regimes that exhibit distinct slopes in the log-log plot. Based on these fits, we categorize the frequency range over which RP changes from 1 to 3 as partially resolving and the RP>3 regime as highly resolving. The fit results are \(\alpha = 2.39 \pm 0.33\) in the partially resolved regime and \(\alpha = 0.9 \pm 0.33\) in the highly resolved regime. In \cref{sec:Resolving Screens}, we demonstrate that for two screens scaling in size proportional to \(\nu^{-2}\), the resolution power \(RP \propto \nu^{-3}\). Substituting this into \cref{Eq:nu_s_for_formula} yields a scintillation bandwidth \(\nu_{\text{s,MW}} \propto \nu^{1}\). Our simulations show that the transition from \(\alpha = 4\) to 1 happens over a range of frequencies and observing only the partially resolving regime can lead to incorrect conclusions. Interestingly, our simulations converge with the expectation from \cref{Eq:nu_s_for_formula}, derived for an incoherently emitting plane, in the highly resolved regime. This suggests that the scatter-broadened source in the first screen retains some level of coherence when partially resolved by the second screen.

\subsubsection{Screen resolving an emission region}
\label{Subsec: Discussion-screen-emissionregion}

\cite{kumar2024constraining} classifies FRB emission models into two categories: nearby models, in which the FRB is generated in a magnetosphere, and far-away models, in which the FRB is generated in shocks at the interface between a wind and surrounding material. The study further argues that the diffractive length, or the lateral size resolvable by a plasma screen in the host galaxy's ISM, lies between the lateral sizes of these two classes of emission models (with the former being $<10^7$ cm and the latter $>10^9$ cm). Based on this, the study proposes that if a FRB originates from shocks, a host screen modulation index \(m_{\text{host}} < 1\) should be observed, resulting from the screen resolving an incoherent emission region.

In the simulations, we model an incoherent emission region as a plane populated with independently emitting points. This approach is motivated by the argument presented by \cite{cordes2016supergiant}, which suggests that what we observe as a single FRB is the result of numerous small, coherent pulses that, when incoherently summed, produce a powerful, short-duration radio burst. In an emission region–screen system, the light from the emission region passes through image points in the screen before reaching the observer. Summing the intensities at the observer from all emission points passing through all image points effectively simulates an incoherent source–screen system. To probe the broadband nature, we create 14 simulations over the range of 0.5 to 5 GHz by fixing the screen sizes in a reference simulation and scaling them to other frequencies according to the \(\nu^{-2}\) dependence of scatter-broadening.

\begin{figure}
    \centering    
    \includegraphics[width=\linewidth]{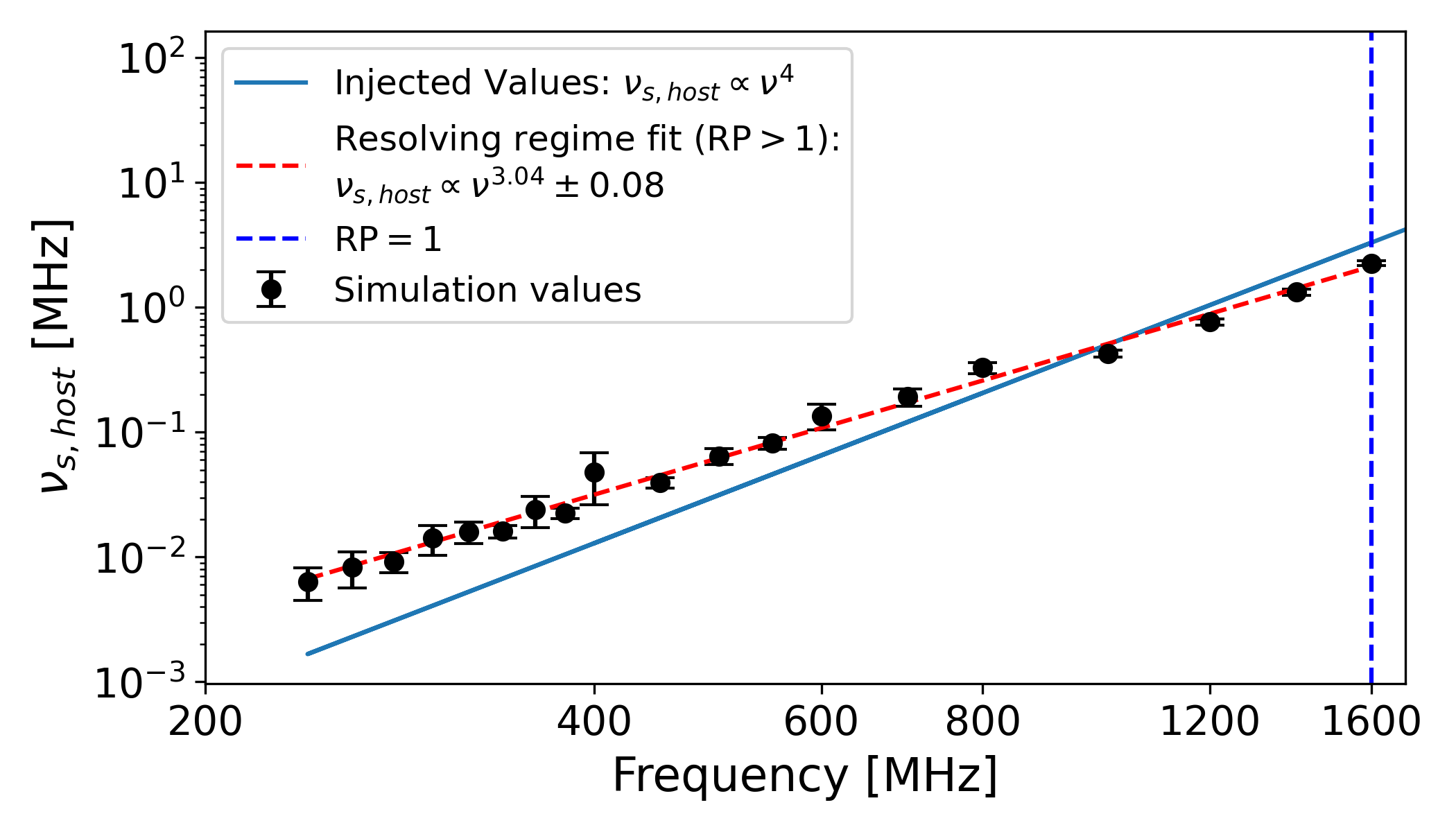}
    \caption{Scintillation bandwidth evolution with frequency (\(\nu_{\text{s,host}} \propto \nu^{\alpha}\)) for a screen resolving an incoherent emission region that is invariant with frequency. The simulations are done in the resolving regime (RP $\in$ [7,1]). Black dots represent mean values from 50 simulations, with error bars indicating the standard error. The red dashed line fits the simulation points, yielding $\alpha = 3$, indicating that the screen resolves the FRB emission region. It deviates from the $\alpha = 4$ trend expected for a point source modulated by a single screen (blue dashed line).}
    \label{fig:scintBW_freq_emissionregion}
\end{figure}

Similar to a two-screen system, the resolution power of a screen–emission region system is also frequency-dependent. For an emission region size that is independent of frequency and a scattering screen size that scales with \(\nu^{-2}\), \cref{eq:RP_def1} reduces to \(\text{RP} \propto \nu^{-1}\). Substituting \(\text{RP} \propto \nu^{-1}\) into \cref{Eq:nu_s_for_formula} predicts that the scintillation bandwidth scales with frequency with a power of \(\alpha = 3\) in the resolving regime, and \cref{fig:scintBW_freq_emissionregion} shows that our simulations agree with this theoretical expectation. 

This phenomenon has been observed in FRB 20221022A and has been used to constrain the emission region to the magnetosphere of the FRB \citep{2025Natur.637...48N}. This result seemingly contradicts the argument in \citet{kumar2024constraining} that FRBs of magnetospheric origin cannot be resolved by the host screen. However, this argument is limited to the case where the host screen whose scintillation is observable lies far away from the magnetosphere. Also, the limit of \(<3\cdot10^4\) km reported by \citep{2025Natur.637...48N} is still at the boundary of the resolvable regime as defined by \citet{kumar2024constraining}. A similar study could also be employed to constrain the radius-to-frequency mapping of radio emissions from neutron stars, such as in pulsar emissions and giant pulses.
\section{Summary and Outlook}

The primary aim of this work was to explore the impact of propagation through two scattering screens on an FRB, motivated by many FRBs showing observational evidence of scattering and scintillation on two different scales. Two scattering screens can either spatially resolve each other or not resolve each other depending on the physical size of the screens, their distance, and the observing frequency. These two regimes leave different scattering and scintillation signatures on the FRB, and exploring these observational signatures is the crux of the paper.  First, a theoretical framework to study multi-screen scattering in fast transient sources incorporating the cosmological nature of screen distances was presented. The theoretical predictions were then verified using a new simulation tool to study two-screen scattering in FRBs, {\tt FRB\_scintillator}. In the simulations, we placed the two screens in the MW and FRB host galaxy, to be concrete, and explored the resolution effect into three regimes: unresolving, just resolving, and completely resolving.

Even among the earliest FRB discoveries, it was pointed out that the presence of MW scintillation required the screens to be unresolved \citep{masui2015dense}, and this allowed several authors to place constraints on the distance between the FRB source and host galaxy screen \citep{masui2015dense,Farah2018MNRAS.478.1209F,Ocker2021ApJ...911..102O,main2022scintillation,sammons2023two}. The standard argument was that a screen resolving another screen case is analogous to a screen resolving an incoherent emission region, thereby quenching all scintillations. Instead, we show that only the broad-scale scintillations are quenched, but the more correct geometric interpretation are ``mixing" terms in the geometric delay due to the propagation between the screens are prominent in the completely resolving regime. 

To summarize the main results of this work:
    \begin{enumerate}
    \item  The resolving effects in a two-screen system quench only the broader scale scintillations of the two scales produced by two screens, regardless of which screen encounters the source first. From this we conclude that when one screen resolves the other then the reverse is also always true.

    \item \textbf{Spectral ACF and Modulation Index:} The spectral ACF of a burst encountering $n$ screens along the line of sight that are point like to each other has the form \(\text{ACF}_{\text{tot}} = \prod_i( 1 + \text{ACF}_{\text{i}}) - 1\). Correspondingly the total modulation index has the form \(m = \sqrt{2^n - 1}\). As the screens resolve each other the relatively broad scale scintillations are further broadened and quenched. The extent of quenching depends on resolving regime of the system. In the unresolving regime, the total two-screen modulation is \(m_{\text{tot}}=\sqrt{3}\), and the broad-scale scintillation has a modulation of \(m_{\text{broad}}=1\). When the screens partially resolve each other \(m_{\text{tot}}<\sqrt{3}\) and  \(m_{\text{broad}}<1\). In the completely resolved regime, the broad scintillation completely disappears, and only the narrow scintillation is left behind giving an \(m_{\text{tot}}=1\).
    
    \item \textbf{Two-screen trends in broadband studies:} Resolving increases the bandwidth of the broad-scale scintillation and can be identified in multi-frequency observations by studying the frequency dependence of scintillation bandwidth, \(\nu_{\text{s}} \propto \nu^{\alpha}\). In the unresolved regime, \(\alpha \approx 4\), and as the frequency decreases, the two-screen system may transition to the partially resolved regime and then to the completely resolved regime, where the curve flattens to \(\alpha \approx 1\). 

    \item \textbf{Intra-pulse evolution of scintillation (modulation):} 
    For a resolved screen system with an observable scattering tail, the increase in RP over the pulse duration causes increase(decrease) of scintillation bandwidth (modulation) over the pulse duration (\cref{fig:Case3_scint evol} \& \ref{fig: MOD-EVOL}).
    
    \item \textbf{Locating the host screen:} 
    We derive an updated formula (Eq.~\ref{Sachin 2024 formula}) to measure the distance of a host galaxy screen from the FRB with a precision determined by our knowledge of the MW screen location. Once the screen is located, the screen size can be calculated using \cref{Eq:delay_formula}, opening a new avenue of research using FRBs to probe AU-scale extragalactic ISM/CGM structures. If the screen resides in the immediate local environment of the FRB, probing the screen and its velocities offers a way to study the dynamic environment proposed for some FRBs \citep{2020ApJ...899L..21S,main2022scintillation}.
    \item \textbf{1D (elongated) screens:} If screens are spatially elogated, which is commonly observed in pulsars, we show that the alignment of the screens plays a major role in determining how much the screens resolve each other. The orientation of 1D screens also introduces an important caveat when using the formula proposed to calculate the host galaxy screen distance from the FRB in this work (Eq.~\ref{Sachin 2024 formula}) and in previous studies. Any orientation that is not parallel will under-predict the distance between the host screen and the FRB. 

    \item \textbf{Resolving an emission region:} A scattering screen could also, in principle, resolve the emission region in a source of FRBs \citep{kumar2024constraining}. 
    Simulations show that when a screen resolves an incoherent emission region that is invariant with frequency, the scaling of scintillation bandwidth changes from \(\alpha = 4\) to \(\alpha = 3\), as observed in \citealp{2025Natur.637...48N}. To place constraints on the emission region size, one needs to locate the host screen. With just a single pulse, the best-case scenario is to have two screens along the FRB's line of sight that resolve each other, helping to locate the host galaxy screen.
    \end{enumerate}

The theoretical study presented here was performed in light of recent efforts to use observables in two-screen scattering of FRBs to obtain information on source properties. However, we note that future applications of our results may also lie in the modeling of pulsar scintillation. \cref{eq:RP_def1} can be expressed in terms of typical values of pulsar scintillation:
\begin{equation}
    RP = 750 \frac{(1\text{ m})}{\lambda_{\text{obs}}} \frac{L_1}{(1\text{ AU})} \frac{L_2}{(1\text{ AU})} \frac{(1\text{ kpc})}{D_{1,2}}
    \label{eq:RP_def2}
\end{equation}
The screen sizes are denoted by \(L_1\) and \(L_2\), and the distance between them is represented by \(D_{\text{1,2}}\). This formula suggests that if galactic screens are 2D and isotropic, they should completely resolve each other in the strong scattering regime. A system of more than two anisotropic (1D) screens will also inevitably be affected by resolution. We conjecture that such effects could lead to effective isotropic screens that are the result of many screens resolving each other, which would explain that more distant pulsars seem to have more isotropic scattering screens \citep{Oswald/2021MNRAS.504.1115O} as well as the dominance of a single screen that is often observed. The interaction of many screens may also explain the fuzzy secondary spectra observed in pulsars at low frequencies, where the modulation index remains close to one \citep{Lofarcensus12022A&A...663A.116W}. In contrast, in the weak scattering regime, multiple screens can persist without completely resolving each other, as observed in sources like PSR J0437$-$4715 \citep{2025NatAs.tmp..102R} and PSR B1929+10 \citep{2024MNRAS.527.7568O}.

This work emphasizes the advantages of broadband scintillation studies for FRBs and pulsars because of the characteristic spectral evolution of all observables. Ideal instruments for follow-up studies of known repeating sources include the broadband receivers, for example the UBB (1.3–6 GHz) on the Effelsberg 100-m or UWL (0.7–4 GHz) on the Murriyang (Parkes) radio telescopes or other current generation broadband receivers. 

A serious limitation to studies of scintillation scales is the measurement of these scales from an ACF that is not ony affected by instrumental limitations but also by the stochastic nature of the distribution of scattered images on screens, as our simulations have shown. Future research will aim at improved analytical descriptions of the ACF as well as improved parameter inference beyond fits of Lorentzian functions, such that expected stochastic variations are taken into account properly.

In conclusion, resolution effects are an integral part of scintillation and introduce interactions between screens that alter their appearance from established one-screen models.

\begin{acknowledgements}

We thank Kenzie Nimmo, Ziggy Pleunis, Ue-Li Pen, Dylan Jow, Stella Ocker and Mawson Sammons for useful discussions at different stages of this project. We thank the anonymous referee for their very helpful comments. LGS is a Lise Meitner independent research group leader. SPET and LGS acknowledge funding from the Max Planck Society.
\end{acknowledgements}

\begin{appendix}

\section{Doppler rate in a cosmological two-screen system}
\label{sec:fD_2scr}

The Doppler rate is defined as
\begin{equation}
    f_\text{D} =  \frac{\nu \,\der \tau}{\der t} \, .
\end{equation}
This short-time evolution of scintillation can typically only be observed for highly active repeating FRBs \citep{main2022scintillation,wu2024scintillation} which are not treated in this paper. The corresponding derivations are given here for future applications.

Velocities are expressed as physical velocities at the time of the ray crossing the screen, which means angular velocities are redshifted:
\begin{equation}
    \frac{\der\bm{\theta}_{n}}{\der t} = \frac{\bm{V}_{n}}{D_{n} (1 + z_{n})}
    \label{Eq:theta-velocity}
\end{equation}
A compact formulation for the Doppler rate can be obtained by making the further approximation that distances within galaxies are negligible when compared to the distances between galaxies, i.e.~ $D_{\text{src}} \approx D_{\text{MW,host}} \approx D_{\text{host}}$:
\begin{equation}
\begin{split}
    f_\text{D} = - \frac{\nu}{c} \Bigg[ & \frac{D_{\text{host}}\bm{V}_{\text{MW}} - D_{\text{MW}}\bm{V}_{\text{host}}/(1+z) }{D_{\text{MW,host}}} \left( \bm{\theta}_{\text{host}} - \bm{\theta}_{\text{MW}} \right) \\
    & - \frac{D_{\text{src}} \bm{V}_{\text{host}}}{D_{\text{host,src}}} \bm{\theta}_{\text{host}} + \bm{V}_\text{obs} \,\bm{\theta}_{\text{MW}} \Bigg] \, .
\end{split}
\label{Eq:Doppler_formula}
\end{equation}
Here, $\bm{V}_{\text{host}}$ is defined as the relative velocity between host screen and source in order to make the formulation even more compact.

Analogously to the delay, the Doppler rate can be described by defining velocities, while setting the initial positions of observer and source to the origins of their respective planes:
\begin{equation}
    f_\text{D} = - \frac{\nu}{c} \bm{V}_{\text{eff,MW}}\cdot\bm{\theta}_{\text{MW}} - \frac{\nu}{c} \bm{V}_{\text{eff,host}}\cdot\bm{\theta}_{\text{host}} \, . \label{Eq:fD_eff}
\end{equation}
This formulation is phenomenological since it absorbs degenerate parameters into independent degrees of freedom. It is also closely following the canonical formulation of the one-screen case. The effective velocities are given by
\begin{subequations}
\begin{align}
    \bm{V}_{\text{eff,MW}} &= \bm{V}_\text{obs} - \frac{D_{\text{host}}}{D_{\text{MW,host}}}\bm{V}_{\text{MW}} + \frac{D_{\text{MW}}}{(1+z) D_{\text{MW,host}}}\bm{V}_{\text{host}} \, , \\
    \bm{V}_{\text{eff,host}} &= \frac{D_{\text{host}}}{D_{\text{MW,host}}}\bm{V}_{\text{MW}} - \left( \frac{D_{\text{MW}}}{(1+z) D_{\text{MW,host}}} + \frac{D_{\text{src}}}{D_{\text{host,src}}} \right) \bm{V}_{\text{host}} \, .
\end{align}
\end{subequations}

\section{Parameters}
\label{sec: Simulation parameters}

The parameters used for the simulations of two-screen systems in \cref{Sec: Simulation Results} are listed in \cref{tab:parameters_two-screen}. 

\begin{table*}
    \centering
    \caption{Parameters used in simulations in \cref{Sec: Simulation Results}.}
    \label{tab:parameters_two-screen}
    \label{tab:RP=0.2 screen param}
    \label{tab:RP=1 screen param}
    \label{tab:RP=10 screen param}
    \begin{tabular}{c|ccc}
    \hline\hline
    parameter & completely resolved & partially resolved & unresolved \\
    \hline
    $D_{\text{src}}$ & 138 Mpc & 684 Mpc & 684 Mpc\\ 
    $D_{\text{MW}}$  &  1.29 kpc &  1.29 kpc & 2.3 kpc \\
    $D_{\text{host,src}}$ & 3 kpc & 4 kpc & 2 kpc \\ 
    $L_{\text{MW}}$ & 4.12 AU & 4.12 AU & 3.5 AU \\ 
    $L_{\text{host}}$ & 165 AU & 82.5 AU & 20 AU \\ 
    $\tau_{\text{s,MW}}$ / $\nu_{\text{s,MW}}$ & 1 $\mu$sec / 160 kHz & 1 $\mu$sec / 160 kHz & 0.4 $\mu$sec / 0.39 MHz\\ 
     $\tau_{\text{s,host}}$ / $\nu_{\text{s,host}}$  & 0.71 ms / 224 Hz & 0.15 ms / 1 kHz & 19 ms / 8.3 kHz \\ 
    RP & 9.54 & 0.96 & 0.2\\ 
    Intrinsic pulse width (for dynamic spectra) & 0.1 ms & 0.1 ms & 5 ms \\
    \(\nu\) & 800 MHz & 800 MHz & 800 MHz \\
    BW & 25 MHz & 25 MHz & 25 MHz \\
    \hline
    \end{tabular}
\end{table*}

The simulations used for \cref{fig:Dist lim comp} were performed on a band of $500$\,MHz width centered around $1400$\,MHz separated into $2^{13}$ channels. Both the MW screen and the host screen were populated with $1000$ images. The distance to the FRB was determined from the redshift, using a flat $\Lambda$CDM model with $H_0 = 67.37\,\text{km/s/Mpc}$ and $\Omega_{m0} = 0.315$ \citep{Planck2020A&A...641A...6P}. The remaining parameters were drawn from uniform distributions where $0\le z_\text{FRB}\le0.5$, $0\le D_x \le 100$\,kpc, $0\le D_{ys} \le 100$\,kpc, $0\le L_\text{MW} \le 500$\,AU, and $0\le \text{RP} \le 15$. The size $L_\text{host}$ was computed from the other random parameters. Finally, parameter combinations were rejected which did not fulfil $100\nu_\text{s,MW}<\text{BW}$, $\nu_\text{s,MW}>30\delta\nu$, and $\nu_\text{s,MW}>8\nu_\text{s,host}$ where $\delta\nu$ is the channel width. This is done to assert a sufficient number of scintles that are resolved in frequency and to assert that the scintillation scales on both screens are well separated. 87460 individual simulations were performed of which 78494 had convergent fit results. 

The parameters needed to estimate the distance between FRB source and host screen in \cref{sec: host screen distance} are listed in \cref{tab:FRB_parameters}. We confirmed that these numbers lead to the same estimated values as reported in the corresponding paper if the corresponding estimation formula is applied. For \citet{2022ApJ...931...87O} the reference frequency was set to 1\,GHz while for the other studies the center of the observing band was used. For FRB 20190520B, the modulation index is not known and estimations were computed for a value of 1. \citet{2025Natur.637...48N} investigated different orders of the screens. Here, we chose the model where the broader scintillation comes from the MW screen because we focused only on this case.

\begin{table*}
    \centering
    \caption{Measured and estimated parameters for Fast Radio Burst sources used in \cref{tab:formula comparision}.}
    \label{tab:FRB_parameters}
    \begin{tabular}{cc|cccccc}
    \hline\hline
    FRB & Reference & $z_\text{FRB}$ & $\nu$ & $D_x$ & $\nu_{\text{s,MW}}$ & $\tau_{\text{s,h}}$ & $m_{\text{MW}}$ \\
    \hline
    FRB 20190520B & \citet{2022ApJ...931...87O} & 0.241 & 1000\,MHz & 1\,kpc & 52\,kHz & 49.4\,ms & 1? \\
    FRB 20201124A & \citet{sammons2023two} & 0.098 & 713.9\,MHz & 0.46\,kpc & 136\,kHz & 4.04\,ms & 0.59 \\
    FRB 20221022A & \citet{2025Natur.637...48N} & 0.0149 & 600\,MHz & 0.64\,kpc & 124\,kHz & 0.265\,ms & 0.78 \\
    \hline
    \end{tabular}
\end{table*}
    
\end{appendix}

\bibliographystyle{aa} 
\bibliography{references} 

\end{document}